\newif\iffull
\fulltrue  

\iffull
\documentclass[runningheads]{llncs}
\fi

\iffull

\usepackage[table, svgnames,dvipsnames]{xcolor}
\usepackage{wrapfig}
\fi

\usepackage{cite}
\usepackage[english]{babel}
\usepackage{multirow}
\usepackage{url}

\usepackage{soul}
\usepackage{booktabs}
\usepackage[utf8]{inputenc}
\usepackage{color}
\usepackage{eurosym}
\definecolor{dkgreen}{rgb}{0,0.6,0}
\definecolor{gray}{rgb}{0.5,0.5,0.5}
\definecolor{mauve}{rgb}{0.58,0,0.82}
\usepackage{listings}
\usepackage{xspace}
\usepackage{capt-of}
\usepackage[users=BAD]{uchanges}
\ChangesSetUserColor[Bernardo]{B}{red}
\ChangesSetUserColor[Aravinda]{A}{blue}
\ChangesSetUserColor[Aditya]{D}{orange}
\ChangesSetUserColor[Aniket]{AK}{green}
\ChangesSetUserColor[Daniel]{DT}{yellow}
\usepackage{amssymb,amstext,enumerate}
\usepackage{array}
\usepackage[algoruled,linesnumbered,vlined,noend]{algorithm2e}
\SetNlSty{text}{}{:}
	\SetKwComment{Comment}{$\triangleright$\ }{}

\usepackage{graphicx}
\usepackage{listing}
\usepackage{mathtools}
\usepackage{adjustbox}
\usepackage{float}
\usepackage{framed}
\usepackage{tikz}
\usetikzlibrary{arrows.meta,decorations.pathmorphing,decorations.footprints,fadings,calc,trees,mindmap,shadows,decorations.text,patterns,positioning,shapes,matrix,fit}
\usepackage{paralist,balance}
\usepackage{xspace}
\usepackage{float}
\floatstyle{boxed}
\restylefloat{figure}

\renewcommand{\tablename}{Figure}

\iffull \usepackage{hyperref} \fi
\usepackage{bm}
\iffull \hypersetup{
    colorlinks = true,
    citebordercolor = {lime},
    linkbordercolor = {purple},
    urlbordercolor  = {magenta},
    citecolor = {teal},
    linkcolor = {purple},
    urlcolor  = {magenta}
}\fi
\definecolor{light-gray}{gray}{0.80}
\usepackage[capitalize,noabbrev]{cleveref}
\usepackage{tikz}
\usepackage{anyfontsize}
\usetikzlibrary{decorations.pathreplacing}

\usetikzlibrary{arrows.meta,decorations.pathmorphing,decorations.footprints,fadings,calc,trees,mindmap,shadows,decorations.text,patterns,positioning,shapes,matrix,fit,shapes.misc}
\usetikzlibrary{shapes.geometric, arrows, fit}

\tikzstyle{block} = [rectangle, minimum width=1cm, minimum height=1cm,text centered, draw=black]
\tikzstyle{arrow} = [thick,->,>=stealth]

\usepackage{tabularx}
\usepackage{enumitem}

\tikzset{%
    box/.style    = {
rectangle split, rectangle split, 
                      rectangle split parts={#1},
                       font            = \sffamily\footnotesize,
                       text width      = 3.5cm, 
                       draw
                        },
                        sbox/.style    = {
rectangle split, rectangle split, 
                      rectangle split parts={#1},
                       font            = \sffamily\footnotesize,
                       text width      = 1.5cm, 
                       draw
                        },
  wbox/.style    = {
rectangle, 
                       font            = \sffamily\footnotesize,
                       text width      = 4cm, 
                       yshift=2.5em,
                       draw,
                       dashed
                        },
 }

\renewcommand{\paragraph}[1]{\vspace{0.5mm}\noindent \textbf{#1.\ }}

\iffull
\fi

\mathchardef\mhyphen="2D

\newcommand{\NN}{\mathbb{N}}

\newcommand{\spar}{\kappa}

\newcommand{\bit}{\{0,1\}}

\newcommand{\advA}{\mathcal{A}}

\newcommand{\ctr}{\mathit{ctr}}
\newcommand{\round}{r}
\newcommand{\chain}{\mathcal{C}}
\newcommand{\head}[1]{\mathsf{Head}(#1)}

\newcommand{\generalblocki}[1]{\langle \header_{#1},\parsex{#1}{}\rangle (\accset_{#1})}

\newcommand{\Tail}{\ensuremath{\mathsf{Rdb}}}

\newcommand{\transition}{\ensuremath{\delta}}

\newcommand{\ethcodehash}{\ensuremath{\account.h}\xspace}
\newcommand{\ethbal}{\ensuremath{\account.bal}\xspace}
\newcommand{\ethstorageroot}{\ensuremath{\account.sr}\xspace}
\newcommand{\ethnonce}{\ensuremath{\account.nonce}\xspace}
\newcommand{\txto}{\texttt{Tx.to}\xspace}
\newcommand{\txdata}{\texttt{Tx.data}\xspace}
\newcommand{\checkpow}{\ensuremath{\mathsf{chk\_pow}}}

\newcommand{\checkpos}{\ensuremath{\mathsf{vfy\_pos}}}
\newcommand{\solvepos}{\ensuremath{\mathsf{prf\_pos}}}

\newcommand{\length}[1]{\mathsf{len}(#1)}

\newcommand{\prune}[2]{#1^{\lceil #2}}
\newcommand{\pruneback}[2]{{}^{#2\rceil}#1}
\newcommand{\pruneclose}[2]{#1^{#2\rceil}}

\newcommand{\parsex}[2]{ \txset_{#1}^{#2}}
\newcommand{\edit}{{\mathsf{Adb}}}

\newcommand{\policy}{\mathbb{P}}

\newcommand{\redactTx}{\texttt{repairTx}}
\newcommand{\slot}{\mathit{sl}}
\newcommand{\voteTx}{\texttt{voteTx}}
\newcommand{\editchain}{\ensuremath{\mathsf{repairChain}}}

\newcommand{\account}{\ensuremath{\mathsf{Acc}}}
\newcommand{\reqaddr}{\ensuremath{\texttt{REQ\_ADDR}}\xspace}
\newcommand{\voteaddr}{\ensuremath{\texttt{VOTE\_ADDR}}}

\newcommand{\txset}{\ensuremath{\mathit{TX}}}
\newcommand{\accset}{\ensuremath{\mathit{ACC}}\xspace}

\newcommand{\parentheader}{\ensuremath{\mathit{pt}}}
\newcommand{\stateroot}{\ensuremath{\mathit{st}}}
\newcommand{\txroot}{\ensuremath{\mathit{tx}}}

\newcommand{\redactreq}{\mathsf{proposeRepair}}
\newcommand{\vote}{\mathsf{v}}

\newcommand{\votefunc}{\mathsf{Vt}}

\newcommand{\rcbpool}{\mathsf{propPool}}

\newcommand{\genesis}{\mathsf{genesis}}
\newcommand{\voting}{\mathtt{voting}}
\newcommand{\accept}{\mathtt{approve}}
\newcommand{\reject}{\mathtt{reject}}
\newcommand{\candidateblk}{\prop^\star}

\newcommand{\id}{\mathsf{ID}}
\newcommand{\Gtxroot}{\mathit{G}_{\txroot}}
\newcommand{\Gstroot}{\mathit{G}_{\stateroot}}

\newcommand{\checkApproval}{\mathsf{chkApproval}}

\newcommand{\Blk}{\ensuremath{\Gamma}}

\newcommand{\chainValid}{\ensuremath{\mathsf{validateChain}}}

\newcommand{\blockValid}{\ensuremath{\mathsf{validateBlock}}}

\newcommand{\digest}{\ensuremath{\mathsf{retainAndRedact}}}

\newcommand{\isStbl}{\ensuremath{\mathsf{stable}}}
\newcommand{\broadcast}{\ensuremath{\mathsf{broadcast}}}
\newcommand{\tx}{\texttt{Tx}\xspace}

\newcommand{\data}{\mathit{x}\xspace}

\newcommand{\layer}{\ensuremath{\mathsf{Reparo}}\xspace}

\newcommand{\decisionp}{\mathsf{decision}}

\newcommand{\prop}{\mathit{rp}}
\newcommand{\sprop}{\mathit{sp}}

\newcommand{\pf}{\mathit{pf}}
\newcommand{\redactwit}{\mathit{w}}

\newcommand{\pt}{\mathit{pt}}
\newcommand{\hdata}{\mathit{hd}}

\newcommand{\state}{\mathit{st}}

\newcommand{\header}{\mathsf{header}}

\newcommand{\geth}{\texttt{geth}~}
\newcommand{\editTx}{\texttt{repairTx}~}

\iffull
\title{\textsf{Reparo}: \textsc {Publicly Verifiable Layer to Repair Blockchains}}
\titlerunning{Reparo}
\author{%
Sri Aravinda Krishnan Thyagarajan\inst{1} \and 
Adithya Bhat\inst{2} \and 
Bernardo Magri\inst{3,4} \and 
Daniel Tschudi\inst{5} \and 
Aniket Kate\inst{2} 
}
\institute{
  Friedrich-Alexander-Universit\"at Erlangen-N\"urnberg, Germany\and
  {Purdue University, USA} \and
  {Concordium Blockchain Research Center}\and
  {Aarhus University, Denmark}\and
  Concordium, Zurich, Switzerland
}
\authorrunning{Thyagarajan et al.}
\fi

\begin{document}

\maketitle

\iffull
\thispagestyle{empty}

\begin{abstract}
Although blockchains aim for immutability as their core feature,
    several instances have exposed the harms with perfect immutability. The permanence of illicit content inserted in Bitcoin poses a challenge to law enforcement agencies like Interpol, and millions of dollars were lost in buggy smart contracts in Ethereum. A line of research then spawned on redactable blockchains with the aim of solving the problem of redacting illicit contents from both permissioned and permissionless blockchains. However, all the existing proposals follow the build-new-chain approach for redactions, and {\em cannot} be integrated with existing running blockchains, such as Bitcoin and Ethereum.
    
    We present \layer, a generic protocol that acts as a publicly verifiable \emph{layer} on top of  any blockchain to perform \emph{repairs}, ranging from fixing buggy contracts to removing illicit contents from the chain. \layer\ follows the layer design, that facilitates additional functionalities for blockchains while maintaining the same provable security guarantee; thus, \layer can be integrated with existing blockchains and start performing repairs in pre-existing data on the chain. Any system user may propose a repair and a deliberation process ensues resulting in a decision that complies with the repair policy of the chain and is publicly verifiable.
    Our \layer\ layer can be easily tailored to different consensus requirements, does not require heavy cryptographic machinery and can, therefore, be efficiently instantiated in any permissioned or permissionless setting. We demonstrate it by giving efficient instantiations of \layer\ on top of Ethereum (with PoS and PoW), Bitcoin, and Cardano. Moreover, we evaluate \layer\ with Ethereum mainnet and show that the cost of fixing several prominent smart contract bugs is almost negligible. For instance, the cost of repairing the prominent \emph{Parity Multisig wallet} bug with \layer is as low as $0.00005\%$ of the Ethers that can be retrieved after the fix.
\end{abstract}

\newpage
\clearpage
\else

\fi


\section{Introduction}

Blockchain as the underlying technology of cryptocurrencies, such as
Bitcoin~\cite{nakamoto2008bitcoin} and Ethereum~\cite{yellowpaper}
is an append-only,
decentralized ledger equipped with public verifiability and immutability.
While immutability in blockchains 
was always considered attractive, it does come with several issues.
Immutability in monetary aspects is quite unforgiving; e.g., the
infamous DAO attack~\cite{dao} exploited a re-entrance bug in a smart contract
resulting in the loss of $3.6$ million ETH. In Ethereum alone, other
than the DAO bug\footnotemark\ more than $750$K ETH worth more than
\$150 million~\cite{marketcap} (at the time of writing) have been either locked,
lost or stolen by malicious attackers or bugs in smart
contracts~\cite{etherlost,etherlost2,USENIX:BDTJ18}. In a cryptocurrency with
a fixed supply of tokens, stolen or locked tokens pose a huge
problem of deflation~\cite{EPRINT:DDMMT18}, and even worse, could adversely affect the consensus
process on systems based on Proof of Stake (PoS), which Ethereum~2.0 plans to
adopt~\cite{buterin2017casper,casper}. Moreover, writing bug-free software, and
therefore smart contracts, seems to be a long-standing hard problem and the
situation only worsens when many such buggy contracts are uploaded onto the
chain resulting in the loss of hundreds of millions of dollars.

\footnotetext{ The DAO bug was fixed in July 16' by introducing an ad-hoc fix
that runs DAO transactions differently; resulting in a hard fork, that gave
birth to Ethereum Classic. }

Even much-restricted systems such as Bitcoin suffers from the problem of arbitrary data being
inserted in the chain through special transactions,\footnote{Arbitrary
information is permitted in Bitcoin through $\mathtt{OP\_RETURN}$ code, that can
store up to 80 bytes of arbitrary data on the blockchain.} where all miners are
required to store and broadcast the data for validation purposes. Several
academic and law enforcement groups have studied the problem of illicit content
insertion in
Bitcoin~\cite{interpol2015,tziakouris2018cryptocurrencies,matzutt2018quantitative}.
A malicious user can pay a small fee to post illegal and/or harmful content
onto the blockchain via these special transactions.
Interpol~\cite{interpol2015} reported the existence of such arbitrary content
in the form of illicit materials like child pornography, copyrighted material,
sensitive information, etc. on the Bitcoin blockchain. While screening the
contents of a transaction before adding it to the blockchain seems to be a
straightforward solution, Matzutt et al.\ \cite{matzutt2018quantitative} showed
the infeasibility of this approach while giving a quantitative analysis of
already existing contents in the Bitcoin blockchain. Law enforcement
agencies~\cite{tziakouris2018cryptocurrencies} are finding it challenging to
deal with this problem, 

The new General Data Protection Regulation (GDPR) in the European
states has thrown the spotlight on the immutability of \emph{personal
information} like addresses, transaction values, and
timestamps~\cite{ibanez2018blockchains}. These issues could adversely affect the
adaptability of existing blockchain-based applications, especially for
cryptocurrencies if they want to be a credible alternative for fiat currencies.

\subsection{Existing Solutions And Their Limitations}

\paragraph{Redactable Blockchains} The seminal work of Ateniese et al.\ \cite{ateniese2017redactable} was the first
to consider the mutability of blockchains. Their redactable Blockchain protocol
aims to redact illicit contents from a blockchain using chameleon hash
links~\cite{NDSS:KraRab00}. However, their protocol requires the miners to run a
Multi-Party Computation (MPC) protocol which can be quite prohibitive in large
permissionless systems like Bitcoin. Moreover, their protocol requires
modifications to the block structure, making it not useful to remove already
existing illicit content in the chain of Bitcoin or release frozen ethers in
Ethereum. We refer to this property as \emph{Repairability of Existing Contents
(REC)}. Puddu, Dmitrienko and Capkun~\cite{puddu2017muchain}'s proposal suffers
from the same problems, and also, presents the control to modify a transaction
by the transaction creator, which is not useful if the creator does not allow
the desired modifications. Derler et al.~\cite{NDSS:DSSS19} solve the above
problem by using attribute-based encryption where the transaction creator lets
anyone with the right policy modify the transaction. While they do not
require any large-scale MPC among the miners, their protocol lacks public
verifiability and requires modifications on how the Merkle roots are computed in
the blocks, hence does not guarantee REC for Bitcoin or Ethereum.
 The recent work of Deuber, Magri and
Thyagarajan~\cite{deuber2019redactable} leverages on-chain voting techniques to
reach an agreement on the redaction of contents, thereby adding public
verifiability to the redactions. However, their protocol also requires
modifications to the block header and therefore does not guarantee REC for
current systems. Tezos~\cite{goodman2014tezos} proposed a PoS protocol that can instantiate any
blockchain but does not guarantee REC. While lacking formal security guarantees, it also lacks efficiency for multiple
updates. Given that all the aforementioned proposals are \emph{build-new-chain} solutions (no REC) and suffer from other issues as discussed, \emph{none} of them are integrable into existing mainstream permissionless
blockchain systems guaranteeing REC\footnotemark~\cite{brian}. \cref{tab:comparison} summarizes the above discussed limitations. For an extended technical discussion and
comparison, we refer the readers to \cref{sec:related_work}.

\footnotetext{{In case of permissioned setting, Ateniese et
al~\cite{ateniese2017redactable}'s proposal has been commercially adopted by a
large consultancy company~\cite{nytimes, ateniese2018rewritable}.}}

\paragraph{Hard Forks} Performing a repair by forking away from a faulty point in the blockchain can lead to a loss of blocks. A hard fork requires miners to update their client software and corresponding mining hardware. 
Every hard fork brings with it an additional consensus rule in order to validate the whole chain. These additional rules demand additional storage and computational capabilities from clients.
Hard forks are ad-hoc: in Ethereum, DAO was deemed to be a big enough bug to fork the chain, whereas Parity Multisig Wallet was not. \cite{etherlost}

\paragraph{Pruning} For repair operations such as redactions or removing old content, there are pruning solutions that locally redact
contents~\cite{pruning}. However, the primary purpose of this method is space optimization and there is no consensus on what can be removed or redacted. Therefore, a newly joining full node is still expected to receive all the information on the chain for thorough validation.


\begin{table}[!t]
\caption{Comparison of our work with that of the existing redaction solutions. A cross for \emph{Repairability of Existing Contents (REC)} means the proposal is not useful to redact or modify already inserted contents in blockchain.}
\label{tab:comparison}
\centering
\setlength{\tabcolsep}{0.3em} 
\def\arraystretch{1.2}
\begin{tabular}{lcccc}
\toprule
\textbf{Proposals} & \textbf{Stateful} & \textbf{System-scale} & \textbf{REC} & \textbf{Public }\\ 
  & \textbf{repairs} & \textbf{MPC} &  & \textbf{verifiability}\\ \midrule

Ateniese et al.\ \cite{ateniese2017redactable}  & \textcolor{red}{$\times$}&
Required & \textcolor{red}{$\times$}& \textcolor{red}{$\times$}\\ 
Puddu et al.\ \cite{puddu2017muchain}  &  \textcolor{red}{$\times$} & Required &
\textcolor{red}{$\times$}& \textcolor{red}{$\times$}\\
Derler et al.\ \cite{NDSS:DSSS19} & \textcolor{red}{$\times$} & Not required &
\textcolor{red}{$\times$} & \textcolor{red}{$\times$}\\
Deuber et al.\ \cite{deuber2019redactable} & \textcolor{red}{$\times$} & Not
required & \textcolor{red}{$\times$}& \textcolor{blue}{\checkmark} \\ 
Tezos \cite{goodman2014tezos} & \textcolor{blue}{\checkmark} & Not required &
\textcolor{red}{$\times$} & \textcolor{blue}{\checkmark} \\
\textbf{This work} - $\layer$ & \textcolor{blue}{\checkmark} & Not required &
\textcolor{blue}{\checkmark} & \textcolor{blue}{\checkmark} \\ \bottomrule
\end{tabular}
\end{table}

\subsection{Our Contributions}

We present $\layer$\footnote{In the Harry Potter universe, '\layer' is the repairing charm that can be used to seamlessly repair broken objects.} (\cref{sec:editing-new}), which is the first
protocol that acts as a \emph{layer} (in the style of the finality \emph{layer} for blockchains from~\cite{cryptoeprint:2019:504}) on top of {\em any} existing secure
blockchain and allows repair operations on its contents. Our protocol aims to provide a solution that can be easily integrated
into virtually any existing blockchain system, and departs from the \emph{build-new-chain} approach in the literature. 
Although
$\layer$ is bound to the underlying consensus requirements (e.g., PoW, PoS, as discussed in \cref{sec:policy_consensus}), it can easily be adapted to any flavor of consensus (include permissioned systems) without any overhead.
We formally prove
that such an integration of $\layer$ with a secure
blockchain satisfies the standard security properties of \emph{chain quality},
\emph{chain growth} and \emph{editable common prefix} (which were introduced
in~\cite{deuber2019redactable}). We argue that our $\layer$ protocol 
potentially could improve the parameters of the chain quality of the integrated system. 


The main advantage of $\layer$ as a repair layer in contrast to a repairable
blockchain is that, after integration into systems like Bitcoin and Ethereum,
the contents that already exists in these chains become
repairable and thus guaranteeing the REC property: once $\layer$ is
incorporated into the clients of Bitcoin or Ethereum, any previously existing
contents can be repaired. In this direction, we give a detailed instantiation of our $\layer$ integration into
Ethereum when the underlying consensus is PoS (\cref{sec:ethereum_pos}) and PoW (\cref{sec:ethereum}), Bitcoin (\cref{subsec:bitcoin}) and Cardano
(a PoS based system~\cref{subsec:cardano}). We demonstrate how to 
perform repair operations in Ethereum, which can fix smart
contract bugs. For Bitcoin, we show how to perform redaction of arbitrary data
entries (illicit data) that are non-monetary information without changing the
Bitcoin block structure or using any heavy cryptographic machinery. With respect
to~\cite{deuber2019redactable}, our instantiation with $\layer$ has comparable
efficiency in terms of time and is significantly better in terms of space
efficiency: unlike~\cite{deuber2019redactable}, $\layer$ does not require any
additional hash value to be stored in every block header (as detailed
in~\cref{subsec:bitcoin}). Also, $\layer$ is better than a hard fork.
 For instance, consider a situation where a user accidentally
creates a contract with no code in it, it is safe for the user to create a
repair transaction with $\layer$ that adds code to this contract without forking. Users of the
system can skip the expensive, cumbersome and often times arbitrary\footnotemark\ procedures
involving a hard fork.

\footnotetext{ For example, in Ethereum, DAO was deemed to be an important
enough attack requiring a hard fork, whereas the Parity Multisig wallet was not.
}

 

Finally, we offer a proof-of-concept implementation of our $\layer$ protocol
integrated into Ethereum. {As we show in~\cref{sec:implementation}, when
importing the latest $20$ thousand block sub-chain from the Ethereum main network, our baseline implementation has an
overhead of just $0.98\%$ when compared to its vanilla counterpart.} The choice of Ethereum is motivated by the wide-spread adaption and generality of Ethereum's functionalities.

\paragraph{Practical Implications} Apart from illicit data redaction in Bitcoin, for systems such as Ethereum, a repair involves re-running all the transactions that are affected thus demanding computation from the network. Therefore, a repair proposal must pay (in gas) an amount proportional to the computation spent by the network performing the repair. We measure the repair costs of various existing bugs affecting Ethereum today.

{For concreteness, we demonstrate that the Parity Multi Sig Wallet Bug, which locked over $513$K ETH, can be repaired today by paying a little over $0.27$ ETH in gas.} $\layer$ also gives a mechanism to resolve an issue where users submit a contract creation transaction with no code~\cite{EIP156} (due to user errors or buggy wallet code), releasing over $6.53$K ETH.\@ Ethereum uses an ad-hoc fix for DAO as it hard-codes a different logic for DAO.\@ $\layer$ can be used to remove this ad-hoc fix by first repairing DAO code (while the fix is still active) and later removing the fix. One could also handle zero-day vulnerabilities and thereby restrict losses.

Future adopters of blockchains such as governments and corporations can use $\layer$ as a provably secure protocol for regulations and maintenance.




%

\floatstyle{plain}
\restylefloat{figure}
\begin{figure}
    \centering
    \includegraphics[width=0.70\linewidth]{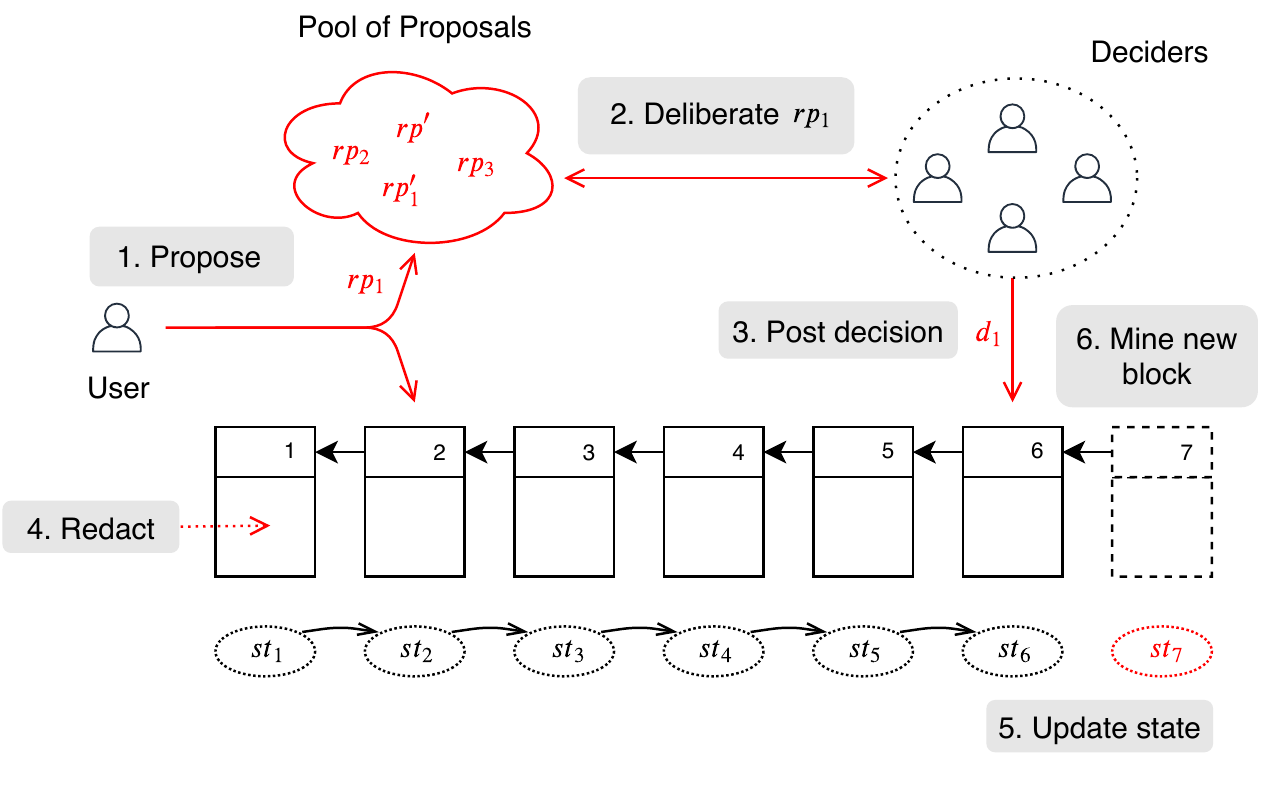}
    \caption{Step by step description of $\layer$ in case of performing a
    redaction. The steps are in gray boxes and highlighted with red color. Here
a block is partitioned into a block header and block body and the block
header stores a pointer to the previous block's header. State $\state_i$ denotes
the state of the chain after block $i$ has been mined.}
    \label{fig:overview_intro}
\end{figure}

\subsection{Challenges}

Before, we present our solution, we present the challenges involved in creating
a protocol that allows repairs to the blockchain state. For the sake of
simplicity, for our discussion here we will refer to a blockchain as a chain
consisting of hash links and state links. Hash links are fields in blocks which
contain the hash of the previous block. Similarly, state links are fields in
blocks which contain the state after applying the transactions in the current
block.

Consider a naive repair protocol which (i) collects votes from the miners (ii)
when sufficient votes from miners within a pre-specified block range, the
proposal is considered approved, and (iii) performs repairs and creates a new
hash and state link chain and joins it at the point of approval.


As observed, this solution is not scalable to multiple repairs without modifying
the contents of the block. The hash links and state links become complex
branched chains. For multiple edits and multiple redactions at arbitrary point,
we will need to store a large branched chain.  This leads to performance losses
and increased synchronization times for new nodes. In our work, we present an
elegant solution with optimizations that reduce these overheads.

\subsection{Key Ideas} $\layer$ acts as a layer on top of an existing
blockchain system. The system users agree beforehand what the policy is going to be
for performing repair operations on the chain. This policy specifies the constraints
and requirements a proposal needs to satisfy for getting approved. As shown
in~\cref{fig:overview_intro}, $\layer$ layer constitutes $6$ sequential steps for
the case of a redaction proposal. Any user can propose a repair proposal
$\prop_1$ for block $B_1$ that redacts some contents in $B_1$. The proposal also determines the updated state of the chain unlike previous proposals. A group of users called
\emph{deciders} deliberate on the proposal (step-$2$) and after the deliberation process, they post their
decision $d_1$ on the chain (step-$3$). After the decision is posted for $\prop_1$, it is
removed from the pool of proposals. Miners check if decision $d_1$ for the proposal $\prop_1$ 
is positive and if $\prop_1$ adheres to the repair policy
guidelines. If so, the miners redact the said contents in block $B_1$ as per the
proposal in step~$4$. Miners update the state of the chain as per $\prop_1$ (step-$5$) and propose a new block $B_7$ with this updated state resulting in an extension of a repaired chain (step-$6$).

The key novelty in $\layer$ is in the consensus-based and cost-efficient (compared to recreating an alternate chain) updating of the state $\state_7$ that accounts for the
states and contents of previous blocks, new incoming contents and the redaction
operation that was performed in step~$4$. $\layer$ also allows repair operations other than redactions also,
and in that case, the miners skip step~$4$ and only update (or repair) the state of
the chain according to the proposal. This state update takes into account the
necessary repairs to the chain that the user wishes to perform.

\subsection{Discussion}\label{sec:remarks}
 Notice that $\layer$ possesses public verifiability of proposals, deliberation and repair operations: $\layer$ has accountability during and after
a deliberation process is over for any repair proposal, referred to as \emph{voting phase accountability} and \emph{victim and new user accountability}~\cite{deuber2019redactable}. In this section we argue about some crucial features of $\layer$ that makes it stand apart from the rest of the proposals.


\noindent\textbf{What if users decide to retain redacted data?} Similar to previous proposals, $\layer$ does not enforce complete removal of redacted data from a user's local storage. Users can still locally keep redacted data, however, once repaired by $\layer$ the users are not required by the blockchain protocol to store the redacted information. For instance, in the case of illicit content, this means that the miners who locally keep and broadcast illicit (redacted) data can be prosecuted individually if necessary and the system as a whole is not liable.\\

\noindent\textbf{Can a bad set of deciders retroactively censor transactions?} Similar to censorship of transaction inclusion by miners, it is also possible to ``censor'' transactions retroactively through repair operations. However, this can be easily mitigated by requiring multiple decider sets across different deliberation phases to approve a repair operation. Thus, a single bad set of deciders at a given time interval cannot censor. Moreover, contrary to the censorship on transaction inclusion, attempts to censor through repair operations are publicly verifiable as the transaction is already on chain and the network is aware of the deliberation process.\\

\noindent\textbf{How is $\layer$ different from the DAO fix in Ethereum?} The hard fork in Ethereum to fix the DAO bug was an ad hoc software patch in the Ethereum client. On the other hand, $\layer$ is a layer on top of the underlying blockchain system that can handle virtually any kind of repair operations subject to restrictions of the policy. \\

\noindent\textbf{Using $\layer$ to perform monetary changes in the state can cause inconsistencies?} Although $\layer$ here is described in a generic way, in Bitcoin for example the repair policy could restrict repair operations to be only redaction of auxiliary data that does not affect user's balances. For Ethereum, the policy could allow contract bug fixes that indeed affects monetary balances of user accounts. \\

\restylefloat{figure}
\floatstyle{boxed}

\section{Blockchain Formalism}

\floatstyle{plain}
\restylefloat{figure}
\begin{table}[!ht]
\caption{Interfaces provided by the blockchain protocol $\Blk$. As an example of the stability interface, in Bitcoin the stable part is the chain pruned by the most recent $k$ blocks (e.g. $k=6$). The idea is that the stable part of a chain will (with overwhelming probability) remain immutable.}
\label{fig:interfaces}
\centering
\def\arraystretch{1.2}
\begin{tabular}{p{.29\columnwidth} p{0.63\columnwidth}}
\toprule
\textbf{Interface} & \textbf{Description} \\ \midrule
$\Blk.\chainValid(\chain)$\ & returns $1$ iff all the blocks in the chain are valid according to a public set of rules and the links between blocks are well-formed\\ \midrule
$\Blk.\blockValid(B)$\ & for a block $B:= \langle \header, \data \rangle$ returns $1$ iff the block is valid; specifically the hash of $\data$ is equal to the data-pointer $G(\data)$ in $\header$.
\\ \midrule
$\Blk.\broadcast(\data)$ & broadcasts data $\data$ to all the parties. \\
  \midrule
$\Blk.\isStbl(\chain)$ & returns the stable part of the chain $\chain$. \\
\bottomrule
\end{tabular}

\end{table}


\subsection{Blockchain as a Ledger}\label{subsec:prelims}
A blockchain is simply a chain (or sequence) of blocks that we call $\chain$. The $i$-th block in the chain $\chain$ is denoted by $B_i := \langle \header_i, \data_i \rangle$, where $\header_i = (\pt,G(x_i), \hdata)$ and $\data_i$ denotes the messages contained in the block. Here, $H:\bit^*\rightarrow\bit^\spar$ and $G:\bit^*\rightarrow\bit^\spar$ are cryptographic hash functions, $\pt$ is the hash of the previous block header $H(\header_{i-1})$ and $\hdata$ contains some special header data, such as the consensus proof $\pi_i$ for the block (e.g., a nonce for PoW or a stakeholder signature for PoS). For a block $B_i$ to be considered valid, it needs to satisfy a public set of requirements established by the protocol; the requirements can vary depending on the application of the blockchain, but at the very least the consensus proof $\pi_i$, and the set of transactions contained in $\data_i$ needs to be valid according to some pre-determined rules. 

The rightmost block is called the head of the chain, denoted by
$\head{\chain}$. The sequence of blocks till the $\head{\chain}$ defines the state $\state$ of the chain $\chain$. Any chain $\chain$ with a head $\head{\chain} :=\langle \header, \data \rangle$ can be
extended to a new longer chain $\chain':=\chain||B'$ by attaching a (valid)
block $B' := \langle \header', \data' \rangle$ where $\pt = H(\header)$ and the state of $\chain'$ is updated to $\state'$; the head of the new
chain $\chain'$ is $\head{\chain'} := B'$. A chain $\chain$ can also be empty,
and in such a case we let $\chain := \varepsilon$. For a chain $\chain$ of length $n$ and any $q \ge 0$, we denote by
$\prune{\chain}{q}$ the chain resulting from removing the $q$ rightmost blocks
of $\chain$, $\pruneclose{\chain}{q}$ to denote the chain with the first $q$ blocks of $\chain$ and we denote by $\pruneback{\chain}{q}$ the chain
resulting in removing the $q$ \emph{leftmost} blocks of $\chain$; note that if $q \ge n$ (where $\length{\chain} = n$)
then $\prune{\chain}{q} :=\varepsilon$ and $\pruneback{\chain}{q}:=\varepsilon$. If
$\chain$ is a prefix of $\chain'$ we write $\chain \prec \chain'$. For a more detailed and precise definition of blockchain and its functionalities and assumptions we refer the reader to~\cite{C:BMTZ17}.

A secure blockchain satisfies the properties of common prefix, chain growth and chain quality~\cite{EC:GarKiaLeo15,EC:PassSS17,kiayias2017ouroboros}. it is shown that these properties of the blockchain satisfy \emph{persistence} and \emph{liveness} with which we can classify it as a ``healthy" blockchain. Intuitively, in a healthy blockchain after some time period, all honest users of the system will have a consistent view of the chain, and transactions posted by honest users will eventually be included. For formal definition of the above properties, we refer the reader to~\cref{sec:security_def}.

%
%
%
%
%

\subsection{Blockchain Protocol}\label{sec:abstraction}


The basis of our $\layer$ protocol is a healthy immutable blockchain protocol
(e.g.,~\cite{EC:GarKiaLeo15, EC:DGKR18}), denoted by $\Blk$ with interfaces as
described in~\cref{fig:interfaces}.
In $\Blk$, parties are categorized into different roles (not mutually exclusive), namely \emph{users} and \emph{miners}. Users can send inputs in the form of messages while miners try to extend the blockchain by creating and appending new blocks containing the users' messages. The existing blockchain protocols~\cite{EC:GarKiaLeo15,EC:PassSS17,kiayias2017ouroboros} achieve the security properties stated previously, by assuming that the majority of miners are \emph{honest}: honest miners behave according to the protocol. We make the same assumption. Therefore, in protocol $\Blk$ we also assume that the majority of miners are honest (if one instantiates $\Blk$ with the protocols from~\cite{EC:GarKiaLeo15,EC:PassSS17,kiayias2017ouroboros}).

 We refer as \emph{node} to any party in the system, be it a user or a miner.  Each node locally stores its current chain $\chain$. We assume that nodes automatically update their local chain whenever there is a \emph{better}\footnote{According to the ``best chain rule" of the underlying blockchain system.} valid chain available. We assume that a node has access to the interfaces as described in~\cref{fig:interfaces}.

\section{Reparo Protocol}\label{sec:editing-new}

In this section we show how to extend a given immutable blockchain protocol $\Blk$ (as described in~\cref{sec:abstraction}) into a \emph{repairable} blockchain protocol $\Blk'$ that permits repair operations ranging from \emph{redactions} to \emph{state updates} on the chain. 

\floatstyle{boxed}
\restylefloat{figure}

\begin{figure}[!th]
\small

	\noindent \textbf{Initialisation (miner/decider).} Initialize the \emph{proposal pool}, $\rcbpool \gets \emptyset$. 
	\smallskip
	\smallskip

	\noindent	\textbf{Proposal (user).} To propose a repair: 
	\begin{compactenum}
		 \item  create repair proposal, $\prop = \langle (\pt, \data') ,\sprop \rangle$. 
		\item  broadcast it to the network, $\Blk.\broadcast(\prop)$.
	\end{compactenum}
\smallskip
	\smallskip

	\noindent	\textbf{Update proposal pool (miner/decider).} In periodic intervals:
	\begin{compactenum}
		\item collect all valid proposals $\prop$ and set $\rcbpool \gets \rcbpool \cup \{\prop\}$.
		\item if proposal $\prop$ has a corresponding repair witness $\redactwit$ in $\isStbl(\chain)$ set $\rcbpool \gets \rcbpool/\{\prop\}$ to remove $\prop$ from the pool.
	\end{compactenum}
\smallskip
	\smallskip


		\noindent \textbf{Deliberation process (decider/miners).} For each new proposal $\prop \in \rcbpool$:
	\begin{compactenum}
		\item deciders deliberate and come to a decision, denoted as $\redactwit \gets \decisionp(\prop)$.
		\item miners add $\hdata \gets \hdata \cup \{\redactwit\}$, where $\hdata$ is part of the next new block.
	\end{compactenum}
\smallskip
	\smallskip


		\noindent \textbf{Repairing the chain (miners).} For each $\redactwit:= \langle \pt,H(\prop), G(\data'),\sprop,b, \pf \rangle \in \isStbl(\chain)$: 

	\begin{compactenum}
	\setdefaultleftmargin{1em}{1.5em}{0.3em}{0.3em}{0.3em}{0.3em}
		\item if $b=1\ \text{and}\ \checkApproval(\policy,\redactwit)=1$,
		\begin{compactenum}
			\item replace the data $\data$ in the block pointed by $\pt$ by the new data $\data'$. If $\data' =\data$ no action is needed.
			\item update state $\state$ of chain $\chain$ to $\state'$ using $\sprop$.
		\end{compactenum}
		\item else if $b=0$, ignore $\prop$ as deciders have rejected. 
	\end{compactenum}	
\smallskip
		\smallskip

	\noindent \textbf{Chain validation (miners).} Update $\Blk.\chainValid$ to handle repair operations: 
	\begin{compactenum}
		\setdefaultleftmargin{1em}{1.5em}{0.3em}{0.3em}{0.3em}{0.3em}
		\item start validating block $B$ from genesis.
		\item for a block $B:= \langle \header, \data \rangle$, where $\header:
			= (\pt,g, \hdata)$, update $\Blk.\blockValid$ to do the following
			checks:
		\begin{compactenum}
			\item if $G(\data)=g$, then no repair has happened, go to
				step~\ref{step:final}.
			\item else, retrieve all repair witnesses of the form $\redactwit :=
				\langle \pt,h', g',\sprop, 1, \pf \rangle \in \chain $ where
				$\pt$ points to $B$.
			\item for each of these repair witnesses $\redactwit$ in the same
				order of their retrieval, do the following steps (exactly as it
				was performed by the miners originally during repairing):
			\begin{compactenum}[(i)]
				\item check if $\checkApproval(\policy,\redactwit)=1$,
				\item check if the corresponding repair proposal $\prop:=\langle
					(\pt, \data') ,\sprop \rangle$ (where $G(\data')=g'$ and
					$H(\prop)=h'$) was performed correctly according to the
					witness $\redactwit$.
				\item check if state updates of $\chain$ was correctly performed
					according to $\sprop$.
			\end{compactenum}
			\item for the final repair witness $\redactwit:= \langle \pt,h',
				g',\sprop, 1, \pf \rangle$ in the order that was retrieved:
			\begin{compactenum}[(i)]
				\item check if it holds that $G(\data)=g'$ to see if the current
					data in $B$ (that is pointed to by $\pt$). This check also
					works for redactions.
				\item check if the final state obtained after applying all the
					state updates $\sprop$ from all witnesses is consistent with
					the state of the chain at $B$ under validation.
			\end{compactenum}
		\end{compactenum}	
		\item\label{step:final} finally, ensure that all repair operations that
			have an approved witness $\redactwit$ on the chain have been
			performed. This check can be performed on the fly as we validate
			blocks from the genesis.
	\end{compactenum}
\smallskip

	\caption{\small Repairable blockchain protocol $\Blk'_\policy$ resulted from adding $\layer$ on top of $\Blk$ and with policy $\policy$.}
	\label{fig:newprotocol}
\end{figure}

\subsection{Reparo Description}\label{sec:editable-newnew}


We add an additional category of parties (apart from users and miners) to the set of parties involved in our protocol, namely \emph{deciders}. Note that the categories are \emph{not} mutually exclusive, e.g., a user can also be a decider and/or a miner.

	The $\layer$ protocol, formally described in~\cref{fig:newprotocol}, allows repair operations on the underlying blockchain $\Blk$: redaction of data from $\Blk$ and/or special changes in the current state of the chain. It  communicates with $\Blk$ through the interfaces described in~\cref{fig:interfaces}. Similar to~\cite{deuber2019redactable}, $\layer$ is parametrized by a repair policy $\policy$. The integration of $\layer$ with the blockchain protocol $\Blk$ is denoted by $\Blk'_\policy$. We describe the two flavors of repairs in more details next. 
	

\noindent \emph{Redactions:} Without loss of generality, we consider a redaction to be the removal of the entire content (i.e., all transactions) of a block. The redactions can be made much more fine-grained (e.g., on the transaction level) by adding cumbersome details, what we defer to~\cref{sec:ethereum_pos}, when we present an instantiation of $\layer$ on top of Ethereum.

 \noindent \emph{Other repair operations:} Or the repair operation acts as any other type of message that changes the current state of the chain from $\state$ to $\state'$. For instance, in this case, in contrast to normal messages such as ``Alice send some coins to Bob" with Alice authenticating this transfer, these repair messages can alter the current balances of Alice and Bob arbitrarily without requiring authentication from either Alice or Bob.

\paragraph{Repair Proposal} Any user can propose any type of repair request. A repair proposal $\prop = \langle (\pt, \data'),\sprop \rangle$ consists of a block-pointer $\pt$ and the new data $\data'$ (or $\bot$ in case of redaction) and $\sprop$ is a (set of) message(s) that describes the desired state change (state $\state$ of $\chain$ is changed to $\state'$). If the repair is a redaction, then the original data $\data$ stored in the block pointed to by $\pt$ is removed, and the state of the chain is updated using $\sprop$. If its not a redaction, then only the state is updated and no data needs to be modified physically (as $\data'=\data$ where $\data$ is stored in the block pointed to by $\pt$). This means that the state change $\sprop$ always describes the transition of state of the chain from $\state$ to $\state'$ irrespective of whether the repair is a redaction or not.

\paragraph{Deliberation} These proposals are collected by miners and deciders. The deciders then use a \emph{publicly verifiable} decision protocol $\decisionp(\prop)$ to deliberate whether a
proposal $\prop$ should be accepted or not. The protocol outputs their final decision in the form of a repair witness $\redactwit$. Miners then add the witness into the header data of the next created block. For concreteness, if we let the decision process follow as in~\cite{deuber2019redactable}, where deciders (i.e., the miners themselves) add their votes to the header data $\hdata$ of their newly created blocks before broadcasting it to the network; the witness can then be easily ``extracted'' from the header data of all the blocks during the deliberation period, by simply counting how many votes supported the proposal.
The repair witness $\redactwit := \langle \pt, H(\prop), G(\data'),\sprop, b, \pf \rangle$ consists of a block-pointer $\pt$, a pointer to the corresponding repair proposal $H(\prop)$ and the pointer to the new data $\data'$, the proposed state-change $\sprop$, the decision bit $b$, and a proof $\pf$ which allows to validate the decision (e.g., verifiable vote count). Note that the proposed state change $\sprop$ could be empty, which is the case when the repair operation is stateless modification.
For security, we require the witness proof $\pf$ to be sound, i.e., it should be infeasible for an adversary to produce a valid proof $\pf'$ for $\prop$ if $\prop$ was \emph{not} accepted by the protocol.


\paragraph{Repair Policy} Repair policy $\policy$ dictates the constraints of different repair operations, e.g., what is the duration of the deliberation period, what type of data can be redacted, what changes in the state are allowed, just as in~\cite{deuber2019redactable}. For our case, as minimum requirements from a \emph{valid} policy $\policy$, we have
\begin{inparaenum}[(i)]
	\item a detailed description of what contents can be redacted and what kind of state changes are allowed,
	\item a well defined period (in rounds) for the deliberation process for each proposal,
	\item the header data $\hdata$ of blocks can not be edited. This implies that repair witnesses of other proposals cannot be edited, and
	\item system parameters that determine block creation are not modified. For instance, one cannot modify the mining difficulty (in case of PoW) that was used in some block in Bitcoin.
\end{inparaenum}

\paragraph{Policy Approval} We assume that there is a predicate $\checkApproval( \policy,\redactwit)$ which determines if a proposal $\prop$ is approved. It takes as input the policy $\policy$, a repair witness $\redactwit:= \langle \pt,H(\prop), G(\data'),\sprop, b, \pf \rangle$ for a repair proposal $\prop$. The predicate outputs $1$ if the proposal $\prop$ is accepted by the deciders $(b=1)$ with a valid proof $\pf$ and complies by the policy $\policy$, and outputs $0$ otherwise. 
For the formal analysis of security we refer the reader to~\cref{sec:security}.

\subsection{Consensus-Specific Repair Policies}\label{sec:policy_consensus}

Here we discuss how the $\layer$ repair policy $\policy$ deals with different consensus specific challenges.

\paragraph{Proof of Work (PoW)} The set of deciders are chosen in a sybil-resistant manner. When the underlying chain is PoW based, one could select the deciders via PoW itself, where the deciders are required to show proof of work. Necessary bounds on the fraction of adversarial deciders are discussed in~\cref{sec:security}. The repair policy of $\layer$ in this setting need not have any restrictions on the kind of repair operations that can be performed: data can be redacted or modified. However, though $\layer$ does not impose any restrictions, some applications may prefer to have policies that allow only restricted repair operations. For instance, if one is interested to redact arbitrary illicit non-payment data from Bitcoin transactions, the $\layer$ repair policy $\policy$ can be set accordingly. On the other hand, if one is interested in fixing buggy contracts that have cost a lot of money and effort (in case of Ethereum), the policy could be set to allow specific repair operations on the state of the system.  

\paragraph{Proof of Stake (PoS)} In case of PoS based consensus, the repair policy should ensure that repair operations do not invalidate consensus. More specifically, the repair policy $\policy$ should disallow redactions of state. This is because PoS inherently relies on the state of the system for consensus, and removing some state information permanently makes the existing consensus proofs unverifiable. Of course, redactions that do not affect the state of the chain can still be performed with $\layer$. The repair policy should also ensure that during the deliberation process the set of deciders do not change. In other words, the deliberation process should happen in a phase where the set of deciders are fixed. This ensures that any repair operation does not affect any other ongoing deliberation process or the decider set.

\section{Instantiation in Ethereum with Proof of Stake}\label{sec:ethereum_pos}
\floatstyle{plain}
\restylefloat{figure}

We discuss PoS in Ethereum and then continue to describe the working of ethereum
today. We then proceed to detail how one can instantiate our $\layer$ layer
protocol of Section~\ref{sec:editable-newnew} on top of Ethereum to support
repair operations: redaction of transaction contents and/or state updates in the
form of smart-contracts ``patches'' and account balance update (e.g.,
restitution of stolen coins).

\subsection{A Primer on Ethereum}

Ethereum~\cite{yellowpaper} is a decentralized virtual machine (Ethereum Virtual
Machine or EVM), which runs user programs - \emph{smart contracts} - upon user's
request. Roughly, a contract is a collection of functions and variables,
where each function is defined by a sequence of bytecode instructions that
operate on the function input and the variables associated with the
contract. The contract has an address for users in the network to interact with,
and this address depends on the contract creator. A user may interact with a
contract through transactions that call functions in the contract.

\paragraph{Transactions and Block Structure} An Ethereum transaction \texttt{tx} can serve two purposes: message calls or special calls. The \texttt{tx.from} field is derived from signature values \texttt{tx.r}, \texttt{tx.s}\footnotemark. The \texttt{tx.to} field contains the $160$-bit address of the recipient. The \texttt{tx.value} field contains the amount of ether (in Wei) to be transferred from the sender to the recipient, and in the case of contract creation it initializes the contract with the amount. The \texttt{tx.data} field optionally contains EVM bytecode for contract creation or an encoding of a function call of a contract. There are special reserved recipient addresses like \texttt{0x00..0-8} for special calls. These addresses contain native contracts. Native contracts contain instructions that are not executed by the EVM. Similar to Bitcoin, Ethereum has a block header and block content associated with a block. The relevant\footnotemark{} contents of the block header are shown and described in~\cref{fig:ethereumheader}.

\addtocounter{footnote}{-1}
\footnotetext{These are ECDSA signatures that help derive the public key and
thus, the sender of the transaction.}
\stepcounter{footnote}

\footnotetext{There are other fields like $\mathtt{ommers\_hash}$, 
$\mathtt{receipts\_root}$, $\mathtt{extra\_data}$ in the Ethereum block header.} 


%
%


\begin{table}[t!]
\caption{Structure of the Ethereum block header. Relating to the abstract
protocol from~\cref{fig:newprotocol}, $G$ is of the form $G =
(\Gtxroot,\Gstroot)$, and $\hdata = (d,t,\ctr)$.}
\label{fig:ethereumheader}
\centering
\def\arraystretch{1.2}
\begin{tabular}{p{.31\columnwidth} p{0.6\columnwidth}}
\toprule
\textbf{Value} & \textbf{Description} \\ \midrule
$\mathtt{parent\_hash}\ (\parentheader)$ & hash\footnotemark of the previous
block header \\ \midrule
$\mathtt{state\_root}\ (\Gstroot)$ & hash of the root node of the state tree, after all transactions are executed and finalizations applied\\ \midrule
$\mathtt{tx\_root}\ (\Gtxroot)$ & hash of the root node of the tree structure populated with all the transactions in the block \\ 
  \bottomrule

\end{tabular}
\end{table}

\footnotetext{Ethereum uses the 256-bit variant of Keccak/SHA3.}


\paragraph{Accounts and State} State in Ethereum is denoted by $\accset$ which
consists of account objects. There are two types of accounts in
Ethereum: the external account and the contract account. Both types of
accounts (\account) contain balance (\ethbal), storage root (\ethstorageroot),
nonce (\ethnonce) and code hash (\ethcodehash). \ethstorageroot is the hash
digest of the trie encoding of the state of the contract while code hash
\ethcodehash is the hash of the contract bytecode. An external account has empty
\ethcodehash and \ethstorageroot. The effect of the transactions included in a
block have on the accounts is the state of the accounts at the time; reflected
in the \ethbal and \ethstorageroot fields of accounts at the time of mining the
block and consequently in the $\mathtt{state\_root}$ stored in the block header. 

The $\accset$ is updated every block by using a global state transition function
$\transition: \bit^* \times \bit^* \rightarrow \bit^*$. This function takes as
input the state of the accounts in the previous block and new transactions
included in the current block and returns the current state of the accounts in
the chain. For block $B_i$ we have  $\accset_i \gets
\transition(\accset_{i-1},\txset_i)$ where $\accset_{i-1}$ is the state of
accounts in $B_{i-1}$. The output of this function can be thought of as the
changing of account states to $\accset_i$ from their previous state
$\accset_{i-1}$ after applying the new incoming transactions $\txset_i$. These
transactions are validated (signature, balance and nonce checks) according to
Ethereum rules before letting them affect the state transition. 
Note that \accset is analogous to UTXO in
Bitcoin and is derived from the block
but does not exist as a part of the chain. 

\subsubsection{Ethereum with PoS}

We briefly describe the Ethereum protocol when its PoW consensus is replaced
with a PoS consensus like Algorand~\cite{gilad2017algorand} or Ouroboros Praos (OP)~\cite{EC:DGKR18}. 
Algorand is a Byzantine fault tolerant (BFT) consensus, where a set of nodes (known as \emph{committee}) is selected through a sortition procedure based on the weight of the stake they
own. Every round, the committee engage in a Byzantine agreement protocol to produce a new block to be appended in the Algorand chain. OP, on the other hand is a slot based consensus protocol where time is divided into slots and blocks are created relative to a slot. Parties with stake can participate in a slot lottery, and winning the lottery (referred to as \emph{slotleader}) allows a stakeholder to create a block in a particular slot. The probability of winning the lottery for a
stakeholder is directly proportional to his stake. Assuming for simplicity that
users in Ethereum have one account each, then OP dictates that the
probability of winning the lottery is : $\phi_f(\account) := 1 -
(1-f)^{{b_\account}/{S}}$, where $\account$ is the account of the stakeholder,
$b_\account$ is the balance in this account, $S$ is the total stake in the
system and $f$ is some difficulty parameter. We make block-box use of the
sortition procedure or the lottery function for our work.

 The main difference in contrast to PoW is that the difficulty $d$ and nonce $\ctr$ are not in the block header. Instead, generating a validity proof for a block is done by the algorithm $\solvepos$ that outputs a proof $\sigma$ on the block with respect to some account (address) referred by $\account$. Verifying the proof of stake is done by the algorithm $\checkpos$.

\subsection{Reparo on Ethereum Protocol}\label{sec:ethereum_adaption_pos}

In this section we describe $\layer$ on Ethereum when the PoS consensus is
instantiated with Algorand or OP. The reasoning is that, in both these
proposals the stakeholder also happen to be slotleaders if they are chosen. 

As in~\cref{sec:editing-new}, we formally denote a block in Ethereum as
$B:=\langle \header, \txset \rangle ( \accset )$, where $\txset$ denotes the set of
transactions (individual transaction is denoted by $\tx$), $\header := (\parentheader ,
\Gtxroot(\txset),\Gstroot(\accset),\hdata)$ as in~\cref{fig:ethereumheader} and
$\accset$ denotes the state of the accounts in Ethereum.
Here $\Gtxroot(\txset)$ is the Merkle root of the transactions,
$\Gstroot(\accset)$ is the Merkle root of the account state.


Regarding the roles of parties when PoS is instantiated with Algorand or OP,
we consider \emph{miners} or \emph{slotleaders} in the PoS setting also to be
\emph{deciders} of repair proposals. This means that the deliberation process
happens on the chain with \emph{slotleaders} voting on proposals by adding
special voting transactions in the header data $\hdata$ of their blocks. Recall
that in Algorand \emph{slotleaders} are referred to as the \emph{committee members} who
are chosen to propose a block at that round.
Due to space constraints, we give a formal description of the protocol in~\cref{sec:ethprotocol_pos_formal} (\cref{fig:ethprotocol_pos}).

On a very high level, while performing repair operations, we repair the block
contents using new $\layer$ data structures. This ensures that the
block header always remains unchanged while only the block contents are
repaired. This enables efficient multiple repairs on a block: multiple repairs
on the same block's state (direct) or the block's state gets updated multiple
times due to a cascading effect (indirect). Physically repairing block contents
while making use of the $\layer$ data structures also improves efficiency in
terms of consensus for PoS. This is because the blocks always contain the most
recent state of accounts and balances which makes the retrieval of updated
stakeholder distribution in case of PoS easier.

\begin{figure*}[!t]
\centering
\includegraphics[width=0.8\linewidth]{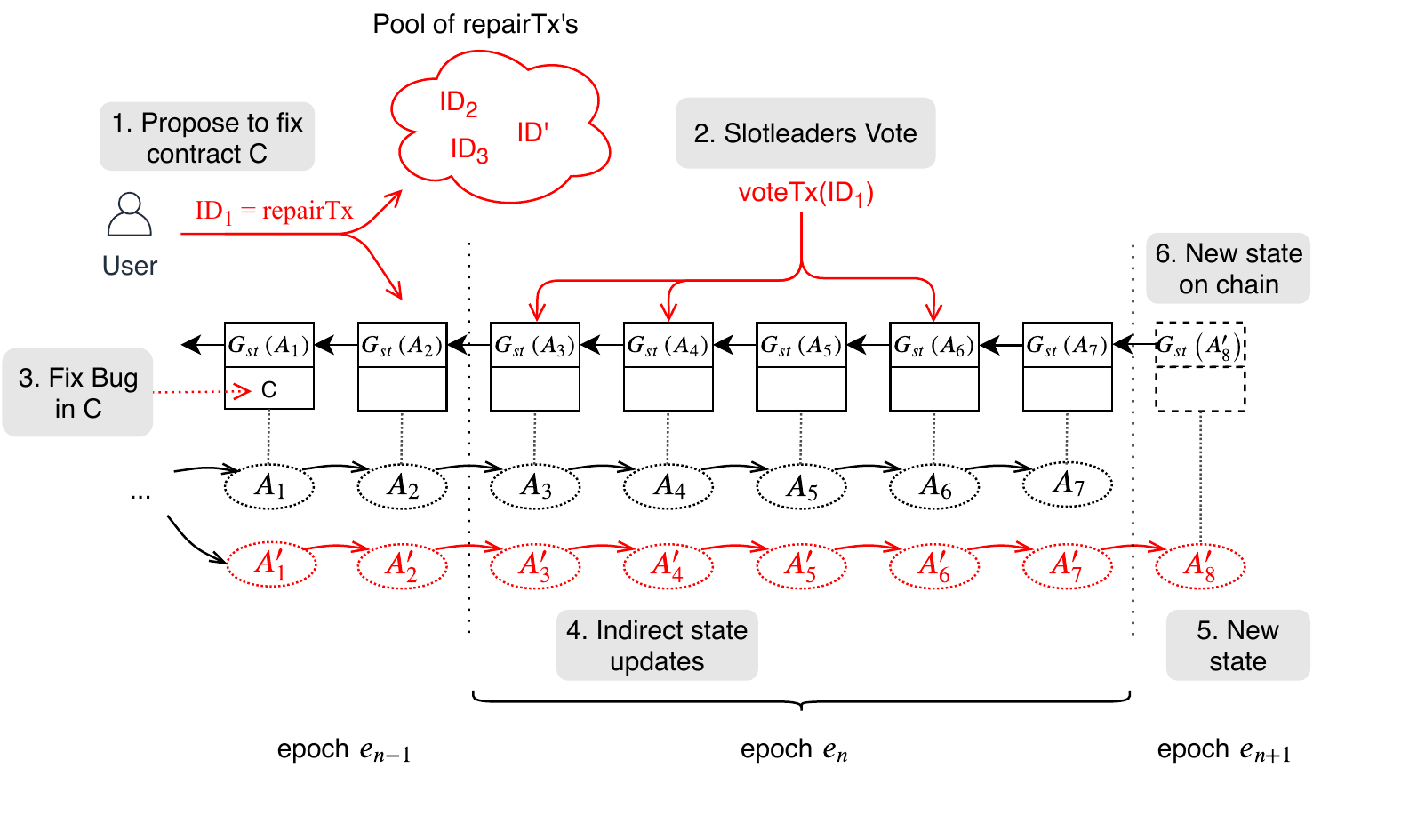}\vspace{-5ex}
\caption{\small An overview of $\layer$ in Ethereum with PoS to fix a buggy contract $C$
in block $1$. The $\layer$ layer steps are numbered inside gray boxes and
highlighted in red. The voting period starts at the start of the next epoch and
lasts for $\ell=5$ blocks. Proposal $\text{ID}_1$ is approved at block $7$. The
voting period coincides with the start and end of an epoch.}
\label{fig:eth_overview_pos}
\end{figure*}

To see how repair operations could affect the state of accounts denoted by $A$, changing
contents can affect the state of the concerned accounts, which could
subsequently lead to cascading changes to other accounts. This is pictorially
described in~\cref{fig:eth_overview_pos} where a user proposes to fix a buggy
contract $C$ in step 1. During the voting period of $\ell(=5)$ blocks that
coincides with a PoS epoch, slotleaders vote for the proposal by adding a vote
inside the block that they propose in step 2. The reason for the voting period
and the epoch to coincide was explained in~\cref{sec:policy_consensus}. If
enough votes are obtained and the proposal satisfies some set of policy
guidelines, in step 3 slotleaders fix the contract C according to the proposal.
The states of all accounts in the subsequent blocks are updated amounting to
this fix in step 4 and a new state $A_8'$ is obtained in step 5. In step 6, this
updated state is reflected on the chain by proposing the next block with respect
to this state (by including $\Gstroot(A_8')$ in the header).

\floatstyle{boxed}
\restylefloat{figure}

\paragraph{Proposing Repairs} Any user in the system can request a
repair of the chain. The user first broadcasts the candidate transaction 
$\tx^\star$ to the network. 
Then, the user sends a proposal transaction \texttt{tx}. 
The \texttt{tx.to} address field contains the special address $\reqaddr$.
\reqaddr is a native contract for \layer. The \texttt{tx.data} field contains
$(H(\tx),H(\tx^\star))$, the hash of the old version ($\tx$) and the new version
($\tx^\star$) of the transaction. For smart contract bug fixes, $\tx$ was the
buggy-contract creation transaction, while $\tx^\star$ is a similar transaction
with the bug fixed. The $\redactTx$ offers processing fees to the slotleader who
includes the transaction into the block and could also offer a approval fee to
the slotleader who performs the repair after the policy approval. The user also
adds the new version of the transaction $\tx^\star$ to the candidate database
for the users to validate and consider voting for the proposal\footnotemark. A
formal description of proposing a candidate block is given
in~\cref{alg:request_pos}. The repair proposal $\candidateblk_j$ can be seen as
including the candidate transaction and other unchanged transactions of that
block (allowing fine-grained transaction level changes). It also includes the
new state of accounts: as the new set of transactions (candidate transaction
plus other old transactions) are applied on the state of accounts of the
previous block. It is interesting to note here that a repair proposal can be
accompanied by a bounty (in \texttt{tx.value}) incentivizing the network to
approve the repair.

\footnotetext{If a candidate transaction does not have a corresponding
$\redactTx$ in the blockchain then the transaction is not included in the
candidate pool, and it is treated as spam instead.}



\begin{algorithm}[!t]
\small
  \SetKwInOut{Input}{input}\SetKwInOut{Output}{output}
   \Input{Chain $\chain = (B_1,\cdots,B_n)$ of length $n$, an index $j \in [n]$, and new set of transactions $\txset_j^\star$.}
  \Output{A repair proposal $\candidateblk_j$.}
  \BlankLine 
    Parse $B_{j-1} := \generalblocki{j-1}$\;
    Build the repair proposal $\candidateblk_j \gets \langle \txset_j^\star
    \vert \vert \transition(\accset_{j-1},\txset_j^\star)\rangle$\;
    \Return $\candidateblk_j$\;
 \caption{$\redactreq$}
 \label{alg:request_pos}
\end{algorithm}

\floatstyle{plain}
\restylefloat{figure}


\paragraph{Validating Requests} 
Nodes validate a repair proposal by checking if the proposed new $\tx^\star$ is
a well-formed transaction as per rules of Ethereum (correct format, correct
signatures, etc) and $\tx$ is in the chain. They also check if the proposal
$\candidateblk_j$ from~\cref{alg:request_pos} includes the correct state of
accounts after applying $\tx^\star$ and other unchanged transactions of block
$B_j$ on the state of accounts in $B_{j-1}$. Proposals are rejected as redundant
if they are already in the voting phase.



\paragraph{$\layer$ Layer} $\layer$ has new data structures that help store the
block contents (transactions and state of accounts) that enable efficient
multiple repairs and chain validation. We have two such data structures:
\emph{repair layer} $\Tail$ database and \emph{approved repairs} $\edit$
database.  Every block is associated with its own repair layer $\Tail$ database
entry that comes into play when the block contents are repaired (directly or
indirectly). For repairs that are not redactions, $\Tail$ of the block contains
the old version $\tx$ of the transaction that was repaired. In case of
redactions, as discussed in~\cref{sec:policy_consensus}, the policy $\policy$
only permits non-payment data to be redacted. These are data (transaction fields
or transactions themselves) whose changes do
not affect the state of accounts. In this case the hash of the old version of
the transaction $H(\tx)$ is stored in the $\Tail$ of the block. In the formal
description~\cref{fig:ethprotocol_pos} the entire set of old transactions and
state of accounts is stored in the repair layer. We emphasize that this covers
the possibility of many transactions in a block being repaired multiple times.
We note that as a practical optimization, you only need the old version of the
specific transaction to be stored. The \emph{approved repairs} $\edit$ database,
stores the repair proposal that was approved by the policy at that time.  This
data structure plays a crucial role in chain validation. The information stored
in these data structures do not need any special authentication, as they can be
validated using simple hash equality checks: in case of information stored in
$\Tail$, the $\header$ of a block stores the corresponding transaction root and
state root, and in case of $\edit$, the chain stores the hash of the repair
proposal (in the form of votes as explained later).

\paragraph{Repair Policy} We briefly discuss the repair policy $\policy$
 for Ethereum with PoS that determines if a
repair proposal has been approved or not. Although our voting based deliberation
process is similar to the protocol in~\cite{deuber2019redactable}, the
deliberation and corresponding policy in $\layer$ is much more complicated.
Therefore, $\Blk'.\checkApproval$ in~\cref{fig:ethprotocol_pos} for the policy
$\policy$ and a repair proposal returns $\accept,\ \reject$ or $\voting$:
$\accept$ and $\reject$ means that the repair proposal has been approved and
rejected, respectively, and $\voting$ means that the repair proposal is still in
deliberation phase. 

Policy $\policy$ takes in the information from the real world like
user discussions, forum discussions, expert opinions, etc. to see if a
particular repair proposal is good for the chain or not. In any case, we wish to
give a minimum policy requirement for redactions and other repair operations
which can later be updated depending on the application. The objective of this
minimum requirement is to enable miners to detect malicious repair proposals
that aim at double spending or stealing coins. We emphasize that this by no
means is a complete set of requirements to detect such behavior, and in fact it
may be of independent research interest to frame such policies for various
applications. In other words, enforcement of the policy is not done on chain. The minimum requirements from the policy $\policy$ are: \begin{inparaenum}[(1)]
\item The deliberation period for the request began at the start of the epoch
	and ended with the end of the same epoch.
\item The proposal does not propose to modify the address fields or the value
	field of a transaction.
\item The proposal does not redact or modify votes in the chain.
\item The proposal has received more than $\rho$ fraction of votes ($50$\% of
	votes) in the epoch ($\ell$ consecutive blocks which is voting period, and
	can decided by the system) after the corresponding $\redactTx$ is included
	in the chain. And finally,  
\item the proposal is (unambiguously) not a double spend attack attempt (which
	needs information from the real world for confirmation).
\end{inparaenum}

\paragraph{Deliberation by Voting} In the deliberation process, the slotleaders
vote for a repair proposal by generating a voting transaction $\voteTx$ and
including that in the block that they propose. If the votes received is approved
according to $\checkApproval$ with policy $\policy$, slotleaders consider these
votes as the witness $\redactwit$ (as in~\cref{fig:newprotocol}). The
\txto address field of the $\voteTx$ is a special address $\voteaddr$ and
the \txdata field contains the hash of the old transaction and the hash of
the candidate transaction.  Formally, we define the interface: 

 $H\left(\Gtxroot(\txset_j),\Gtxroot(\txset_j^\star)\right) \gets \Blk'.\votefunc(\chain,
        \candidateblk_j)$: takes as input a repair proposal
		$\candidateblk_j$ and outputs the hash value of the \texttt{Tx.data} field
        of the corresponding repair proposal as a vote.

Note that we use $\Gtxroot$ instead of $H$ for $\txset_j$ and $\txset_j^\star$,
as we deal with set of old and new transactions. The reason for using
$H\left(\Gtxroot(\txset_j),\Gtxroot(\txset_j^\star)\right)$ stems from the need
to allow redactions and other repairs to be performed on $\txset_j$ or
$\txset_j^\star$, while later a new user can verify this without providing the
original $\txset_j$ or $\txset_j^\star$ itself.


\paragraph{Performing Repairs} Upon approval with respect to the policy $\policy$:

\begin{asparaitem}
	\item \emph{Redactions}: These operations are restricted by the policy $\policy$ to not
		affect the state of accounts in Ethereum chain. Withstanding this
		restriction, the original transaction is replaced by the candidate
		transaction and the repair layer stores the hash of the old version of
		the transaction. We use $\digest(\txset_j)$ function, which returns the
		hash of the redacted transaction and all other unedited transactions in
		$\txset_j$ in the original form~\cref{alg:editchain_pos}. If a version
		of the transaction is stored in either $\Tail$ or $\edit$, it is
		redacted too ensuring that a redacted
		transaction is not stored even in the $\layer$ layer. 
	 \item \emph{Other repairs}: For other repairs, the old
		 version of the transaction is stored in the repair layer $\Tail$
		 associated with the block. The candidate transaction then replaces the
		 old version in the block. The state of accounts (for this block and the
		 following blocks) is updated according to the repaired
		 transactions. When updating the state for each block, we ensure that
		 the original state of accounts is stored in the corresponding repair
		 layer $\Tail$. Once the state updates reach the head of the chain, the
		 slotleader proposes a new block with this updated state of accounts.
		 \cref{alg:editchain_pos} gives a formal description, where entire set
		 of original transactions and state are stored in the repair layer (thus
		 covering the possibility of multiple repair operations on a block). For
		 improving space efficiency one could store only the old version of the
		 transaction.
\end{asparaitem}

Note that, since a repaired block always contains the most recent state,
performing multiple indirect state updates is efficient as we only apply
the transition function over the block's latest contents during each of the state
updates. 


\begin{algorithm}[!htp]
\small
	\SetKwInOut{Input}{input}\SetKwInOut{Output}{output}
\Input{Chain $\chain = (B_1,\ldots,B_n)$ of length $n$, repair layer $\Tail =
(\Tail_{1},\ldots,\Tail_{n})$, and a repair proposal $\candidateblk_j$.}
\Output{Chain $\chain'$, repair layer $\Tail'$}
\BlankLine
Parse $B_j := \generalblocki{j}$, $\candidateblk_j := \langle\parsex{j}{\star} \rangle$\;
Set $B_j^\star \gets \langle \header_j, \parsex{j}{\star}\rangle$\;
\Comment{\scriptsize{If block is never repaired then store original state.}}
\lIf{$\Tail_j = \emptyset$}{
if $\candidateblk_j$ is a redaction proposal, set $\Tail_j \gets \langle \digest(\txset_j)\vert\vert \accset_j\rangle$, otherwise set $\Tail_j \gets \langle\parsex{j}{}\rangle$}\label{line:redact}
\lElse{Parse $\Tail_j := \parsex{j}{'}$, and 
if $\candidateblk_j$ is a redaction proposal, set $\Tail_j \gets \langle \digest(\txset_j')\vert\vert \accset_j'\rangle$}
Initialize $\chain' \gets \pruneclose{\chain}{j-1}\vert \vert B_j^\star,$ and $\allowbreak \Tail' \gets
\pruneclose{\Tail}{j-1}\vert\vert\Tail_j$\;
\Comment{\scriptsize{Update repair layer of blocks in between}}
\For{$i = j+1$ to $n$}{\label{line:loop_update}
    Initialize $\Tail_j^\star = \emptyset$\;
    Parse $B_i := \generalblocki{i}$\;
    \lIf{$i-1 = j$}{ Set $B_{i-1} = B_j^\star$}
    Parse $B_{i-1} := \generalblocki{i-1}$\;
    Set $\txset_i^\star \gets \txset_i$ and $\accset_i^\star \gets \transition(\accset_{i-1},\txset_i)$\;
    Set the block $B_i^\star \gets \langle\header_i, \parsex{i}{\star},\rangle$\;
    \lIf{$\Tail_i = \emptyset$}{
        $\Tail^\star_i \gets (\parsex{i}{})$
    }
    Set $\Tail' \gets \Tail' \vert\vert \Tail_i^\star$, and $\chain' \gets
    \chain' \vert\vert B_i^\star$\;
}
\Return $\chain',\Tail'$\;
\caption{$\editchain$}
\label{alg:editchain_pos}
\end{algorithm}

\pgfarrowsdeclarecombine{twotriang}{twotriang}
{stealth'}{stealth'}{stealth'}{stealth'}

\paragraph{Block Validation} A formal description of the  procedure can be found
in~\cref{alg:validateblk_pos} which is invoked during the chain validation. The procedure checks if the
transactions included in the block are valid as done currently in Ethereum. It
then checks if the hash link is rightly formed. In case no repair proposal has been approved in
this block, the only remaining checks are to see if the state of accounts in the
block are correct and if the slotleader has produced a valid proof of stake.
If any repair proposals were approved by the policy at this block, the procedure performs these repairs on the
chain while performing the required state updates for the blocks. After all the
approved repairs until this block have been applied, the procedure checks if the
state of accounts in the block under contention are consistent with the updated
states of the previous blocks. Finally, it checks if the header is correctly
formed with the correct stakeholder as the slotleader.   

\begin{algorithm}[!t]
\small
\SetAlgoVlined
\SetKwInOut{Input}{input}\SetKwInOut{Output}{output}
\Input{Chain $\chain = (B_1,\cdots,B_n)$, repair layer $\Tail=
(\Tail_{1},\cdots,\Tail_{n})$, block $B_{n+1}$, repair approved $\edit_{n+1}$.}
\Output{$\{\bot,(\chain',\Tail')\}$}

\BlankLine
Parse $B_{n+1} := \generalblocki{n+1}$, where $\header_{n+1} =
(\parentheader_{n+1}, G(\parsex{n+1}{}),\hdata_{n+1})$\;
Parse $B_n := \generalblocki{n}$, where $\header_n = (\parentheader_n, G(\parsex{n}{}),\hdata_n)$\;
Validate transactions $\txset_{n+1}$, if \emph{invalid} \Return $\bot$\;\label{line:transactions}
\lIf{$\parentheader_{n+1} \ne H(\header_n)$}{\Return $\bot$}\label{line:hashlink}
\lIf{$\edit_{n+1} = \emptyset \land \accset_{n+1} = \transition(\accset_n,\txset_{n+1}) \land
\allowbreak \checkpos\left(\chain,\hdata_{n+1},(\pt_{n+1},G(\parsex{n+1}{}) )\right)=1$}{ Set $\chain' \gets \chain \vert\vert B_{n+1}$, and $\Tail' \gets \Tail \vert \vert
\emptyset$, and \Return $(\chain',\Tail')$}\label{line:normalblock}
\Comment{\scriptsize{Validate Blocks where repairs were approved}}
Initialize $\chain' \gets \chain, \Tail' \gets \Tail$\;
\For{all $\txset_j^\star \in \edit_{n+1} $}{
\Comment{\scriptsize{Perform all the repair operations}}
    Parse $B_j := \generalblocki{j}$, where  $\header_j = (\parentheader_j,
    G(\parsex{j}{'}),\hdata_j)$\;
    \lIf{$\checkApproval\left(\policy, H\left(\Gtxroot(\txset_j),\Gtxroot(\txset_j^\star)\right)\right) \ne
    \accept$}{\Return $\bot$}
    Set $\accset_j^\star := \transition(\accset_{j-1},\txset_j^\star)$\;
    \Comment{\scriptsize{Perform the repairs as originally performed}}
    $\chain', \Tail' \gets \editchain(\chain', \Tail',
    \candidateblk_j )$\;\label{line:editchain}
}
Parse $\chain' := (B_1', \cdots, B_n')$ and $B_n' := \langle \header'_n, \parsex{n}{'} \rangle$\;
\Comment{\scriptsize{Check the state transition after repair}}
\lIf{$\accset_{n+1} = \transition(\accset'_n, \txset_{n+1}) \land \allowbreak 
\checkpos\left(\chain,\hdata_{n+1},(\pt_{n+1},G(\parsex{n+1}{}) )\right)=1$}{Set $\chain' \gets \chain \vert\vert B_{n+1}$, and $\Tail' \gets \Tail \vert \vert
\emptyset$ and \Return $(\chain',\Tail')$} 
\Return $\bot$\;
\caption{$\blockValid$ }	
\label{alg:validateblk_pos}
\end{algorithm}

\paragraph{Chain Validation} On receiving a new chain, the chain validation
procedure formally described in~\cref{alg:validate_pos} starts validating the blocks
from the genesis of the chain. It first switches the block contents with the
corresponding transactions and states stored in the repair layer. This results in
the chain ($\chain_\mathit{org}$) in its originally mined state. The procedure then validates each block
as discussed above using $\Blk'.\blockValid$. This results in performing the
repairs (both redactions and other repair operations) as they were performed in sequence. We then obtain a chain in its updated current state and is checked if it is the one that was received.

\begin{algorithm}[!t]
\small
	\SetAlgoVlined
	\SetKwInOut{Input}{input}\SetKwInOut{Output}{output}
	\Input{Chain $\chain = (B_1,\cdots,B_n)$ of length $n$, repair layer $\Tail =
		(\Tail_{1},\cdots,\Tail_{n})$ and approved repairs $\edit = (\edit_1, \edit_2,\ldots,
		\edit_n)$.}
	\Output{$\bit$}
	
	\BlankLine
    Initialize $ \chain_{org} \gets B_1$\;
	\For{$j=2$ to $n$}{\label{line:chainoriginal}
		Parse $B_j := \generalblocki{j}$, where $\header_j = (\parentheader_j, G(\parsex{j}{'}),\hdata_j)$\;
		Parse $\Tail_j := (\txset^\star||\accset^\star) $\;
		\lIf{$\txset^\star\vert\vert \accset^\star = \emptyset $}{
			$\parsex{j}{\star} \gets \parsex{j}{}$
		}
		\lElse{
			$\txset_j^\star||\accset_j^\star \gets \txset^\star||\accset^\star$
		}
		\Comment{\scriptsize{In case of redactions, we have $\txset^\star =\digest(\txset_j)$, from which the original transaction merkle root $\Gtxroot(\txset^\star)$ can be computed}}
		\lIf{$ \Gtxroot(\txset'_j) \ne \Gtxroot(\txset_j^\star) \lor \Gstroot(\accset'_j) \ne \Gstroot(\accset_j^\star)$}{\Return $0$}
        $\chain_{org} \gets \chain_{org} \vert\vert \langle \header_j,
        \parsex{j}{\star} \rangle$\;
	}
    Initialize $ \Tail^\star \gets \emptyset$\;
	Parse $\chain_{org} := (B_1^{org}, \ldots,B_n^{org} )$\;
	\Comment{\scriptsize{Validate each block starting at genesis}}
	\For{$j = 2$ to $n$}{\label{line:loop_validate}
        Set $\mathit{op} \gets \blockValid(\pruneclose{\chain_{org}}{j-1},
			\Tail^\star,B^{org}_j, \edit_j)$\;
		\lIf{$\mathit{op}= \bot$}{\Return $0$}
        \lElse{Parse $\mathit{op} := (\chain',\Tail')$}
Set $\chain_{org}\gets \chain'\vert\vert \pruneback{\chain_{org}}{j}$,
$\Tail^\star \gets \Tail'$\;
	}
	\lIf{$\chain_{org} = \chain \land \Tail^\star = \Tail$}{
		\Return $1$
	}
	\Return $0$
	\caption{$\chainValid$}	
	\label{alg:validate_pos}
\end{algorithm}

\subsection{On Security and Optimizations}~\label{sec:discussions}

We discuss briefly $\layer$'s security, and other optimizations possible for our Ethereum instantiation.

\noindent\textbf{Security} Since the hash function $H$ is modeled as a random oracle (RO), finding a collision on a vote (which is the hash of the ID of the old transaction and the ID of the candidate transaction) is highly improbable. Therefore, when a slotleader votes for a repair proposal in his block, no adversary can claim a different repair proposal for the same vote value. Similarly no adversary can find a different block that hashes to the same hash of an honestly proposed block. Therefore an adversary cannot break the integrity of the chain. Together, they imply the \emph{unforgeability} of votes: if an adversary wishes to vote, he has to possess enough stake to propose a block with his vote himself. Assuming appropriate threshold on adversarial stakes and the honest stakeholders follow the policy $\policy$, $\layer$ integration satisfies \emph{editable common prefix} and preserves chain quality and chain growth.


\paragraph{Effect on Stake Distribution} $\layer$'s repair operations, like
fixing smart contract bugs, affects the balances of users and hence the stake
distribution is altered. During the deliberation phase, it is the stakeholders
who vote for a request fully aware of how the stake distribution  change affects
them. Assuming rational slotleaders and honest majority in the stakeholders, a
slotleader votes for those repair requests that are not obvious double spend
attacks and has least negative impact on his stake. The honest behaviour is
enforced through public verifiability of a slotleader's votes.

\paragraph{Optimizations} To lower the costs of repair operations in Ethereum, we propose \emph{Depth-based future approval}: Depending on the depth $d$ of
the contract that needs repairs, the system can have a parameter $p$
that integrates the fix into the main chain in block number $d/p$ after
approval. For example, at block number $8$M if a contract deployed at
block number $1$M was found to have a vulnerability, then with $d=7$M,
$p=1000$, the fix will be integrated into the chain $7,000$ blocks after
the corresponding $\editTx$ is approved. This alleviates the
computational load on the network by giving them more time to perform
repairs that are deep in the chain. Few other optimizations are discussed in~\cref{sec:ethprotocol_pos_formal}.

\section{Experiments in Ethereum}\label{sec:implementation}

In this section, we report a proof-of-concept implementation of the $\layer$
protocol on top of Ethereum~\cite{Ether}. 

We implement two new types of transactions, namely $\editTx$ and $\voteTx$, and
measure their performance with respect to a baseline transaction in Ethereum.
{We also measure the overhead of implementing these special transactions on the
Ethereum main network by measuring the time taken to import the \emph{latest}
$20$ thousand blocks.} We measure the time taken to import the blockchain because
these introduce overheads for syncing (fully/partially) with the network (see~\cref{table:overhead comparison}). 

In Ethereum, computation is measured in terms of the gas it needs to run the
transaction in the Ethereum Virtual Machine (EVM). Hence, we take a look at the
gas costs to repair (by fixing) some popular bugs by computing the transaction
dependency graph for the contract creation transaction for these bugs. We
estimate the gas cost to re-run all the dependent transactions and provide
real-world numbers on the cost of such repairs in \cref{table:bug cost}.

\paragraph{Setup and System Configurations} We modify the Go client for Ethereum
(\texttt{geth}) for our experiments. We use the version
\texttt{1.9.0-unstable-2388e425} from the official Github repository as
the base version. We set the \geth cache size to $10,000$ MB and disable the P2P
discovery (using the \texttt{--nodiscover} flag). {The import was done using an
export file consisting of blocks from block number $10,903,208$ to $10,929,312$
(latest block as of Sep 25, 2020)} created by the export command from a fully
synced node.

Our experiments employed the following hardware/software configuration:
\textit{CPU}: 24 core, 64-bit,
\texttt{Intel\textregistered~Xeon\textregistered~Silver 4116 CPU} clocked at
2.10 GHz; \textit{RAM}: 128 GB; \textit{OS}: Ubuntu; \textit{Kernel}:
4.15.0-47-generic.


\paragraph{System-Level Optimizations}
We employ the following system-level optimizations in our implementation.
\begin{asparaenum}
	\item \emph{Database choice for $\layer$}: \geth implements three types
		of key-value databases: \emph{Memory Databases} which reside in the
		system memory, \emph{Cached} and \emph{Uncached Databases} which reside
		on the disk. The repair layer only stores active requests and the votes
		for these requests. Hence, a memory database is ideal to
		implement $\editTx$ and $\voteTx$.
	\item \emph{Native Contracts for }$\editTx$\emph{and }$\voteTx$: Native
		contracts (also referred to as \emph{Pre-compiled} contracts) are client-side implementations of functionalities that are
		too complex or expensive (in terms of gas) to be implemented inside the
		EVM. For example, the Ethereum yellow paper~\cite{yellowpaper} uses
		these native contracts to perform SHA3 and \emph{ecrecover} (a function
		that returns the address from ECDSA signature values $r,s$). We use
		native contracts to support $\layer$.
	\item \emph{Fast sync and light-client friendliness}: Fast sync
		is a mode used by the Ethereum clients. In this mode, the clients
		download the entire chain but only retain the state entries for the
		recent blocks (pruning). In bandwidth, our implementation only
		needs to download $|\chain|+m$ from full nodes, where $m$ is the number
		of updates and the final space storage is still $|\chain|$ as the nodes
		can discard the repair layer after syncing.
\end{asparaenum}

\subsection{Special Transactions: {\normalfont \editTx}, {\normalfont
\voteTx}} 

Our two special transactions $\editTx$, and $\voteTx$, have special \texttt{to}
addresses $\mathtt{REQ\_ADDR = 0x13}$ and $\mathtt{VOTE\_ADDR = 0x14}$
respectively.


The transactions are always collected in the transaction pool. We modify the
transaction pool logic, specifically $\mathsf{validateTx}()$. After ensuring
well-formedness of inputs, for $\editTx$ we check that the
\texttt{data} field is exactly $64$ bytes long and the first $32$ bytes
correspond to the transaction hash of an existing transaction in the chain. For
$\voteTx$, we check that the \texttt{data} field contains exactly $32$ bytes.

The input for $\editTx$ consists of hash of the transaction $H(\tx)$ which is
to be repaired and the hash of the proposed new transaction $H(\tx^\star)$. The
validation logic ensures that $\tx$ exists in the blockchain (repaired
blockchain) by adding a new function $\mathsf{isTransactionTrue}()$. In the
implementation of the native code for this transaction, we add the request to
the request memory database, indexed by $ID=H(H(\tx)\vert\vert H(\tx^\star))$
and initialize it with $0$ votes. This database is created on demand. The
footprint of the database is small as we will need to process about $16,384$
repair requests before occupying $1$ MB. In contrast, the default cache memory
used by the client ranges from $512$ to $4096$ MB depending on the client
version and is therefore a safe assumption to make.

The input for $\voteTx$ is the $ID$ described previously. The validation logic
ensures that the input is well-formed (of correct length). In the implementation
of the native code for this transaction, we check if the request exists in the
request memory database. If found, it increments the vote by one. Otherwise, it
throws an error and aborts the transaction.

To evaluate the performance overheads of the special transactions on the client
(and the network), we compare it with a baseline transfer transaction involving
a transfer of ETH between two accounts. The transfer function has the lowest gas
requirements ($21,000$). $\editTx$ ($5.90\%$ overhead) takes $76.09$ ms and
$\voteTx$ ($0.055\%$ overhead) takes $71.89$ ms when compared to a transfer
transaction which takes $71.85$ ms on an average over $100$ iteration. (Refer
\cref{table:overhead comparison}.)

\begin{table}[b!]
    \centering
    \caption{Comparison of operations between the modified client and the unmodified client}
    \def\arraystretch{1.2}
    \begin{tabular}{l l rr}
      \toprule
        \multirow{2}{*}{\textbf{Operation}} & \multirow{2}{*}{\textbf{Type}} &
        \multicolumn{2}{c}{\textbf{Client Type}}\\ \cmidrule{3-4}
        & & \textbf{Unmodified} & \textbf{Modified}  \\ \midrule
        $\editTx$ 	& Time (ms) 	& - 	  & $\mathbf{76.09}$ \\ 
        $\voteTx$ 	& Time (ms) 	& - 	  & $\mathbf{71.89}$ \\ 
        Transfer 	& Time (ms) 	& $71.85$ & $\mathbf{71.85}$ \\ 
        Import 		& Time (Hours) 	& $2.26$& $\mathbf{2.28}$\\ 
        Import 		& Speed (Mgas/s)& $39.70$ & $\mathbf{39.40}$ \\ 
      \bottomrule
    \end{tabular}
\label{table:overhead comparison}
\end{table}

\subsection{Performing Repairs}

\begin{table}[!htp]
    \centering
	\caption{Estimated repair costs (today) using state assertion in \layer for
      Ethereum-PoW. \emph{K} and
      \emph{M} stand for Kilo ($10^3$) and Mega
      ($10^6$) multipliers respectively.}
    \def\arraystretch{1.2}
	\begin{tabular}{l r rr}
      \toprule
		\multirow{2}{*}{\textbf{Bug}} & \multirow{2}{*}{\textbf{ETH Stuck}} &
		\multicolumn{2}{c}{\textbf{Costs of repair}} \\
		\cmidrule{3-4}
		& & \textbf{Gas} & \textbf{ETH}
		\\ \midrule
		DAO & $3.60$ M & $3.81$ M & $0.53$ \\
		QCX & $67.32$ K & $4.7$ M & $0.65$ \\
		Parity & $517.34$ K & $1.95$ M & $0.27$ \\
		REXmls  & $6.67$ K & $1.69$ M & $0.23$ \\
		No Code & $6.53$ K & $438.85$ M & $0.44$ \\
      \bottomrule
    \end{tabular}
    \label{table:bug cost}
\end{table}



In this series of experiments, we analyze the impact of supporting $\layer$ on
client software. For every block, supporting $\layer$ adds an overhead of
checking for approved repairs. If approvals are found, we repair the block body
accordingly. In this section, we analyze the read-write overheads to support the
repair, the cost of building new states and applying transaction dependencies to
repair some real-world bugs (check \cref{table:bug info} for details about
these bugs). We use an unedited (clean) chain for our experiments.

\paragraph{Read-Write Costs} In this experiment, we measure the time to update
the data of a block. This experiment helps to estimate the I/O overheads of
transaction updates in the blockchain. A repair consists of finding a
transaction in the blockchain and replacing it with a new transaction. The
transaction repair overhead consists of the time taken by a node to read the
transaction metadata (block hash, block number and the transaction index in the
block) and write the new transaction data. We point the old hash to the new
transaction data so that when the hash of the old transaction is accessed, the
repaired transaction is furnished by the blockchain. We measure the read and
write times for $10,000$ random transactions from random blocks in
the chain. Random transactions ensure that internal (database, software or
operating system) caches do not skew the measurements. The time taken to read
the metadata is $649.81\mu $s and the write operation takes $2.32$~ms on average
over $100$ runs for each of the $10,000$ transactions.

\begin{figure}[H]
	\centering
	\includegraphics[width=0.65\linewidth]{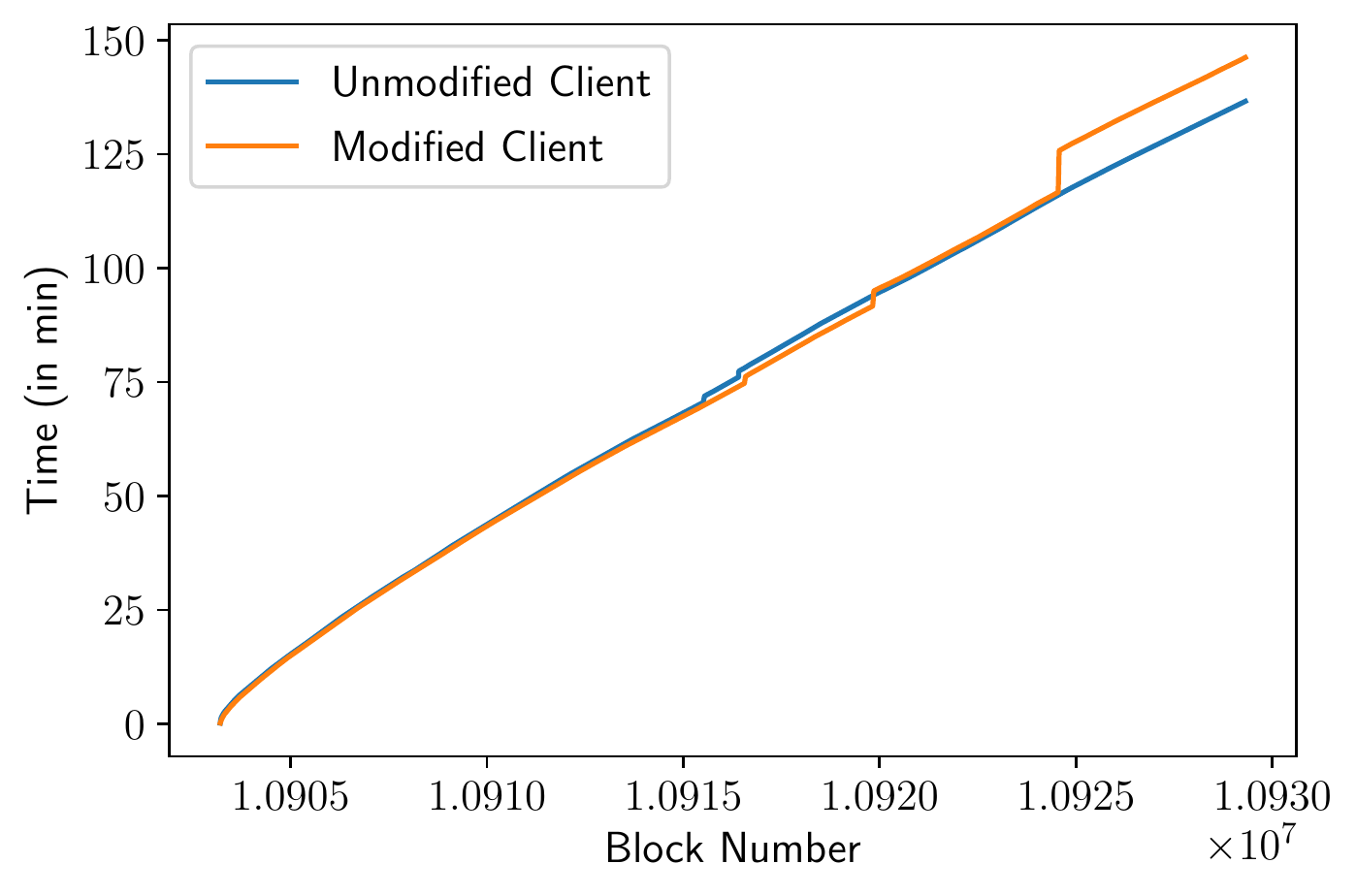}
	\caption{\small Batched speed comparison of the modified client and the
		unmodified client. The modified client has a modified
		\texttt{validateTx} rule, a new function \texttt{isTransactionTrue}, and
		modified structures with flags to detect a dirty (edited chain and
		blocks).  }
	\label{fig:gas-vs-block}
\end{figure}

\paragraph{Import Costs} In this experiment, we evaluate the time it takes to import the Ethereum chain subset using the modified and unmodified versions of the client to measure the impact of \layer in everyday performance. The \geth client imports blocks in batches. We log the amount of gas (in million gas) in such batches and the time elapsed for the import (and thus compute the speed). We perform $3$ iterations on both the modified and unmodified clients.  We plot these speeds for the entire import process for the unmodified and modified clients in~\cref{fig:gas-vs-block}. As evident from the graph, for most of the parts the modified client is equal to or slightly slower than the unmodified client. This is reasonable in the real world as the slight import delay per block can be accounted for by reducing the gas limit of the block (and thus the computation performed on each block allowing $\layer$ to utilize the remaining time).

On average, the unmodified client takes $8134.26$ seconds to import $26,104$ blocks whereas our modified client takes $8213.64$ seconds to import the same blocks. (Refer to \cref{table:overhead comparison}.) This is just $0.98\%$ overhead for a full import of more than $20$ thousand blocks. It does not have any significant effect on the block generation, block validation or block propagation as this can be tweaked by reducing the difficulty and/or gas limit of the blocks.

The average amount of gas processed by the unmodified client is $39.70$ million
gas per second whereas the modified client processes $39.40$ million gas per
second (\cref{table:overhead comparison}). This $0.76\%$ overhead is due to the
hard coding of rules for special transactions whose conditions are checked for
every transaction. This overhead does not cause any problems as the average gas
limit for an Ethereum block is $10,000,000$ (which is under $39$
Mgas/s)~\cite{AvgGasLimit} and both the nodes perform optimally to sync the
latest blocks and propagate. Note that this affects the full sync nodes only.
Note that the light clients, such as Parity~\cite{ParityWarp} for example, skip
verification of states and are thus unaffected.  


\paragraph{Repairs} We employ a policy which allows editing any contract call in order to repair the chain.  We qualify our previous pessimistic analysis by arguing that most of the repairs have small transaction dependency graphs. This is due to the localization of impact to a few accounts.  We bound the number of transactions that need to be re-run to transactions that interact with the contract. This coupled with the fact that we are performing a repair ensures a small transaction dependency graph which significantly reduces the repair costs. \cref{table:bug cost} we highlight the impact of such localizations. We sum the gas in all such transactions to estimate the gas cost of repairs and thus the ETH. Note that we always pay the miners (and hence the network) for the extra computation. We use a gas price of {$138$ GWei/gas (market price at the time of writing)} for our conversions. We refer the interested readers to \cref{sec:eth repair} for more details about the bugs and our solutions.


\section{Conclusion and Future Work} 

This work presents $\layer$, a secure, systematic way to make any blockchain
forget the ``forgettable''. We present a generic protocol that is adaptable to consensus requirements, and achieves public verifiability and secure chain repairs guaranteeing REC for current mainstream blockchains. We then design and analyze an important
application of the protocol in Ethereum to fix contract bugs, and 
report the implications and feasibility of these repairs for popular contract
bugs such as DAO and Parity Multi Sig Wallet. 
We also provide optimizations that can make the implementation more robust and
realizable. We show that, in Ethereum, vulnerabilities, if found, (and existing
vulnerabilities) can be immediately isolated to reduce the transaction
dependency and repaired efficiently and securely.

In the future, we aim to realize the $\layer$ protocol on permissioned systems
such as Hyperledger.
We also intend to study the
impact of $\layer$ on off-chain protocols and whether it can be used to improve
 them.
Among other repair operations, $\layer$ also offers a means to propose, deliberate and incorporate
new features into Bitcoin and Ethereum given the respective communities
currently do this in an ad-hoc manner~\cite{bitBIP,EthEIP}.

\section*{Acknowledgment}
We would like to thank Andrew Miller for his valuable comments and constructive feedback. We would also like to thank all the anonymous reviewers for their insightful comments and suggestions to improve the draft.

The first author was supported by the German research foundation (DFG) through
the collaborative research center 1223, and by the state of Bavaria at the
Nuremberg Campus of Technology (NCT). NCT is a research cooperation between the
Friedrich-Alexander-Universit\"at Erlangen-N\"urnberg (FAU) and the Technische
Hochschule N\"urnberg Georg Simon Ohm (THN). This work also has been partially
supported by the National Science Foundation under grant CNS-1846316.

\iffull
\bibliographystyle{splncs04}
	\bibliography{cryptobib/abbrev1,cryptobib/crypto,cryptobib/extrarefs}
	\appendix

\section{Detailed Related Work}\label{sec:related_work} To solve the problem of
illicit data stored in blockchains, Ateniese et
al.~\cite{ateniese2017redactable} proposed the first redactable blockchain
protocol which uses chameleon hash links~\cite{camenisch2017chameleon}. Their
solution is catered to the permissioned setting where select miners can come
together and redact contents from the blockchain using a large scale MPC
protocol. Unfortunately, their large scale MPC in a dynamic entry-exit
permissionless setting seems to make it infeasible for their proposal to solve
the above discussed problems in Bitcoin and Ethereum. Their proposal apart from
requiring modifications to the block header structure, does not work with SHA256
and requires chameleon hashes. Therefore the protocol is not backward compatible
to any of the existing chains like Bitcoin or Ethereum. 

Puddu et al.~\cite{puddu2017muchain} proposed a protocol where senders encrypt
all but one version of their transactions to the miners along with a
\emph{mutation} policy. The un-encrypted version remains the valid version on
the chain. The miners abiding by the policy can decrypt (via MPC if decryption
keys are shared) alternate versions and make those versions valid. However, a
malicious sender may not include any alternate version at all or may have a
mutation policy where only he can make retrieve the alternate versions.
Moreover, similar to~\cite{ateniese2017redactable}, this proposal too suffers
from scalability issues with large scale MPC in a permissionless setting and is
not backward compatible with existing chains.

Derler et al.~\cite{NDSS:DSSS19} proposed attribute based modification of chain
contents while relying on chameleon hashes. Unlike
in~\cite{ateniese2017redactable}, here chameleon hashes are not used for hash
links but for transaction hashes in computing the merkle root. Any user can tag
an object with an access policy before posting it on the blockchain and only the
users with attributes satisfying this policy can later decide to modify the
object. However, in their setting changing from the old version to the new
version of the transaction does not affect the transaction merkle root.
Therefore a user cannot decide whether and where something was changed or not.
This creates problems when the underlying consensus mechanism is PoS that relies
on the state (i.e., accounts and balances) of the chain. A new user can no
longer verify the consensus that was generated with the old version of the
transaction as it does not exist anymore. In other words, the proposal lacks
accountability as the rewritings are indistinguishable and lacks verifiability
of (state dependent) consensus with respect to new users. Also, similar
to~\cite{puddu2017muchain} this proposal relies on the user to set an attribute
policy which may not be useful if the user sets a policy that only his own or
his colluder's attributes can satisfy. 

Deuber, Magri and Thyagarajan~\cite{deuber2019redactable} proposed the first
efficient redactable protocol for a permissionless setting. They rely on
achieving voting based consensus to perform redactions. The block structure is
modified to have two hash links instead of one and if the previous block is
redacted, one of the links breaks while the other holds. This gives
accountability and public verifiability of where and what was redacted from the
lens of a new user unlike the above mentioned proposals. They do not make use of
any heavy cryptographic machinery and therefore achieve better efficiency in the
permissionless setting. As their focus is only to redact illicit contents that
does not affect payment information, their protocol fails to deal with stateful
edits like doing smart contract bug fixes in Ethereum that have a cascading
effect. Since their protocol is tailor made for PoW based systems, it is unclear
how they do stateless or stateful edits in PoS based systems. And finally, their
protocol is not backward compatible with any existing blockchains given their
requirement of the block structure modification. For a more detailed comparison
with~\cite{deuber2019redactable}, we refer the reader to~\cref{subsec:bitcoin}.

Florian et al.~\cite{florian2019erasing} propose for miners to locally drop
harmful data. Although efficient in case of Bitcoin, they do not focus on global
consensus on what to be erased. Differing miners end up in different forks which
severely limits their functionality, which is aggravated in case of stateful
edit operations. Politou at al.~\cite{politou2019blockchain} present a
comprehensive survey of the various solutions that have been proposed to edit a
blockchain and also give the relevance of GDPR laws for blockchains. 

Tezos~\cite{goodman2014tezos} proposed a generic and self-amending blockchain.
They provide a generic interface for meta-upgrades, i.e changes to the code. The
interface is generic and can be instantiated on any blockchain such as
Bitcoin~\cite{nakamoto2008bitcoin} and Ethereum~\cite{yellowpaper}. Tezos
creates a testnet with the proposed changes/amendments and if there is
sufficient confidence in the testnet (via votes from stakeholders) promotes the
testnet as the main protocol. 

This schemes has several drawbacks. Tezos can instantiate any blockchain
protocol by using the appropriate genesis block.  However, Tezos cannot be
instantiated on an existing blockchain. In other words, Tezos cannot be used to
repair existing blockchains. 

Another drawback of this scheme is that at any point in time, only one proposal
is under consideration (by being in the testnet). The proposal under
consideration is always the one with the most approvals. Even if all the nodes
in the system agree to the change, it takes a minimum of four quarters and two
rounds of voting to integrate the change. This is inefficient when compared to
our proposed scheme.


 \section{Security Definitions}\label{sec:security_def}
 
  The \emph{common-prefix} property states that if one take the chains of two honest users at distinct time slots, the shorter chain (minus a few blocks) is a prefix of the longer chain. This property implies the immutability of the underlying blockchain $\Blk$. \emph{Chain growth} property intuitively says that the chain $\chain$ will eventually grow in number of blocks as the protocol progresses. The \emph{chain quality} property says that the ratio of blocks produced by malicious users in the chain $\chain$ can be upper bounded. 
 
 \begin{definition}[Common Prefix~\cite{EC:GarKiaLeo15}]\label{def:common_prefix}
	The chains $\chain_1,\chain_2$ possessed by two honest parties at the onset of the slots $\slot_1 < \slot_2$ are such that $\prune{\chain_1}{k} \preceq \chain_2$, where $\prune{\chain_1}{k}$ denotes the chain obtained by removing the last $k$ blocks from $\chain_1$, where $k \in \NN$ is the common prefix parameter.
\end{definition}

\begin{definition}[Chain Growth~\cite{EC:GarKiaLeo15}]\label{def:chain_growth}
	Consider the chains $\chain_1,\chain_2$ possessed by two honest parties at the onset of two slots $\slot_1,\slot_2$, with $\slot_2$ at least $s$ slots ahead of $\slot_1$. Then it holds that $\length{\chain_2}-\length{\chain_1} \ge \tau \cdot s$, for $s \in \NN$ and $0<\tau \le 1$, where $\tau$ is the speed coefficient.
\end{definition}

\begin{definition}[Chain Quality~\cite{EC:GarKiaLeo15}]\label{def:chain_quality}
	Consider a portion of length $\ell$-blocks of a chain possessed by an honest party during any given round, for $\ell \in \NN$. Then, the ratio of adversarial blocks in this $\ell$ segment of the chain is at most $\mu$, where $0 <\mu \le 1$ is the chain quality coefficient.		
\end{definition}

\section{Security Analysis}\label{sec:security}
In this section we formally argue the security properties of the repairable blockchain $\Blk'_\policy$ resulting from the composition of an immutable blockchain $\Blk$ and our repair layer $\layer$ in the presence of a \emph{valid} policy $\policy$. By validity of $\policy$ we mean that the policy satisfies the minimum requirements listed above.

Recall that the underlying blockchain $\Blk$ is assumed to satisfy the security properties of \emph{Chain growth}, \emph{Chain quality} and \emph{common prefix}, formally stated in~\cref{sec:security_def}. Also, note that the assumptions of the underlying blockchain $\Blk$ must still hold (e.g., trusted majority), and in particular this means that in a PoS blockchain the majority of the stake must be in the hands of honest users during the entire lifetime of the system. We show that the protocol $\Blk'$ still \emph{preserves} chain growth and chain quality. By preservation of the property we mean that the resulting protocol has \emph{at least} the same guarantees as the original protocol $\Blk$, but potentially stronger. 

\paragraph{Chain Growth} Assuming that $\Blk$ satisfies chain growth it is not hard to see that the $\layer$ added on top of $\Blk$ \emph{does not} influence the chain growth rate of the resulting protocol $\Blk'$, as $\layer$ does not dictate how often new blocks are created and appended to the chain. We give the corollary statement below.


\begin{corollary}
	If $\Blk$ satisfies $(\tau,s)$-chain growth, then $\Blk'_\policy$ preserves $(\tau,s)$-chain growth for any valid policy $\policy$.
\end{corollary}

\paragraph{Chain Quality} Interestingly, when we assume that the majority of the deciders are honest and a majority endorsement is required for $\decisionp$ to output a witness, the $\layer$ protocol can potentially ``improve'' the chain quality coefficient of the resulting repairable blockchain $\Blk'$. To see this, consider an adversarially produced block $B_i \in \chain$. A repair operation $\prop$ proposed and accepted for block $B_i$ could be seen as ``turning'' the block $B_i$ into an \emph{honest} block since the contents of $B_i$ are now agreed by the protocol. This is because for the repair to be performed, it needs to be accepted by $\decisionp(\prop)$ which needs a majority of the deciders (i.e., miners in the case of a permissionless setting) to endorse. Hence, by the honest majority assumption of the underlying blockchain $\Blk$, any accepted repair operation must be backed by \emph{at least} $1$ honest miner (or more, depending on the policy $\policy$), thereby increasing the ratio of honest blocks in $\chain$. 

\begin{corollary}
	For all witnesses $\redactwit_i \in \chain$, let $\pf \in \redactwit_i$ be a sound proof. If $\Blk$ satisfies $(\mu,\ell)$-chain quality, then $\Blk'_\policy$ preserves $(\mu,\ell)$-chain quality for any valid policy $\policy$.  	
\end{corollary}

\paragraph{Common Prefix} It can happen that two honest miners will perform the same repair operation at different times, and in the period in between it can happen that they \emph{do not} have a common prefix. Note however, that a repair is only performed once its accepting repair witness is in the stable part of the chain. Hence, we do not have to deal with rollbacks. There are two observations to be made:
\begin{asparaitem}
\item  This time period is small (i.e., $1$ network delay). If the witness is stable for one miner then it must become stable for the other honest miners as soon as they see all the blocks that the first miner saw. Therefore, the repair operation might briefly disturb common prefix, but not for long.
\item If a repair operation is a redaction that does not alter the state of the chain, the common prefix can be momentarily violated, but at no point it is violated when just considering the state of the chain.
\end{asparaitem}

Even though $\Blk'$ does not satisfy the common-prefix property as stated in Section~\ref{sec:security_def}, following the lines of~\cite{deuber2019redactable}, we show that the protocol $\Blk'$ satisfies the \emph{Editable common prefix} property introduced by~\cite{deuber2019redactable}.

\begin{definition}[Editable Common prefix]\label{def:editable_common_prefix}
	The chains $\chain_1,\chain_2$ of length $l_1$ and $l_2$, respectively, possessed by two honest parties at the onset of the slots $\slot_1 \le \slot_2$ satisfy one of the following:
	\begin{enumerate}
		\item $\prune{\chain_1}{k} \preceq \chain_2$, or
		\item for each $B_j \in \prune{\chain_2}{(l_2-l_1)+k}$ such that $B_j \notin \prune{\chain_1}{k}$, it must be the case that $\exists \redactwit_i \in \prune{\chain_2}{(l_2-l_1)+k}$ such that $\checkApproval(\policy,\redactwit_i) = \accept$ and $\pt_i:=H(B_j)$.
	\end{enumerate}
	Here, $\prune{\chain_2}{(l_2-l_1)+k}$ denotes the chain obtained by pruning the last $(l_2-l_1)+k$ blocks from $\chain_2$, $\policy$ denotes the chain policy, repair witness $\redactwit$ corresponds to a redaction proposal $\prop$, $\pt_i$ is the pointer contained in $\redactwit_i$, and $k \in \NN$ denotes the common prefix parameter.
\end{definition}

\begin{theorem}\label{thm:editable_common_prefix}
	If $\Blk$ satisfies $k$-common prefix, then $\Blk'_\policy$ satisfies $k$-editable common prefix for a valid policy $\policy$.
\end{theorem}

\begin{proof}
	If no repair operations were performed in the chain $\chain$, then the protocol $\Blk'_\policy$ behaves exactly like the protocol $\Blk$.  Henceforth the common prefix property follows directly.
	
	However, in case of some repair operations, consider an adversary $\advA$ that proposes a repair $\prop := \langle (H(\header_i), \data_i'),\sprop \rangle$ to repair contents of $B_i$ in chain $\chain_2$. The proposal is later accepted and the repair witness $\redactwit = \langle H(\header_i), H(\prop),G(\data_i'),\sprop, 1, \pf \rangle$ is included in the chain which is then executed by an honest party $P_2$ at slot $\slot_2$. Observe that by the unforgeability property of the witness proof $\pf$, $\advA$ is not able to efficiently produce a valid proof $\pf'$ for another repair proposal $\hat \prop$ that was not accepted. Therefore, since $P_2$ is honest and incorporated the repair $\prop$ in $\chain_2$, it must be the case that $\prop$ was accepted by \emph{at least} the majority of the deciders. Thus making all the honest parties incorporate the repair $\prop$. This concludes the proof.
\end{proof}

\section{Reparo in Ethereum}\label{sec:ethprotocol_pos_formal}
Formal description of the $\layer$ protocol in Ethereum with a PoS consensus is given in~\cref{fig:ethprotocol_pos}.
\floatstyle{boxed}
\restylefloat{figure}

\begin{figure*}[!htbp]
\smaller
The Ethereum protocol PoS consists of a sequence of rounds (slots) $\round$ and an epoch $e$ consists of $\ell$ rounds. 

\begin{flushleft}
\textbf{Initialization.} We initialize new databases, the repair layer
$\Tail_0 \leftarrow \genesis$, and the approved repairs $\edit_0 \gets \emptyset$, set
round $\round \leftarrow 1$ and an empty list of repair proposals $\rcbpool \gets \emptyset$.
\end{flushleft}
For a given epoch $e$ and for each round $\round$, first initialize $\edit_{r}
\gets \emptyset$, $\Tail_{r} \gets \emptyset$ and we describe the following
sequence of execution.

\begin{flushleft}
\textbf{Proposal.} A node creates a repair proposal $\candidateblk_j
\leftarrow \Blk'.\redactreq(\chain,j, \txset_j^\star)$ (refer~\cref{alg:request_pos}) for block $B_j, j \in [r-1]$ using transactions
$\txset_j^\star$. It then broadcasts it to the network.
\end{flushleft}

\begin{flushleft}
\textbf{Update Proposal pool.} Collect all repair proposals $\candidateblk_j$
from the network and add $\candidateblk_j$ to $\rcbpool$ iff $\candidateblk_j$
is valid; otherwise discard $\candidateblk_j$. If $r$ is the beginning of an epoch, then set $\Blk'.\checkApproval(\policy,\vote_j):=\voting$ where $\vote_j$ is a vote for $\candidateblk_j$.
\end{flushleft}

\begin{flushleft}		
\textbf{Repairing the chain.} For all repair proposals $\candidateblk_j :=
\langle \parsex{j}{\star} \rangle \in \rcbpool$, we denote a vote $\vote_j \gets \Blk'.\votefunc(\chain,\candidateblk_j)$ and do:
	\begin{compactenum}
        \item If $\Blk'.\checkApproval(\policy, \vote_j) = \accept$,
            then call algorithm $(\chain',\Tail')\gets \Blk'.\editchain(\chain,\Tail,\candidateblk_j)$ (refer~\cref{alg:editchain_pos}). Here $j$-th block in $\chain'$ is $\langle
            \header_j,\candidateblk_j \rangle$ and subsequent blocks' states
            are updated accordingly. Then do the following,
         \begin{compactenum}
         	\item Add $\txset_j^\star$ to $\edit_{r}$ and remove $\candidateblk_j$ from $\rcbpool$
         	\item set local chain $\chain = \chain'$ and update $\Tail = \Tail'$
         \end{compactenum}
        \item If $\Blk'.\checkApproval(\policy, \vote_j) =
            \reject$, then remove $\candidateblk_j$ from $\rcbpool$ 
        \item If $\Blk'.\checkApproval(\policy, \vote_j) = \voting$, then do nothing 
	\end{compactenum}
\end{flushleft}
	
\begin{flushleft}
	\label{step:deliberation_pos}\textbf{Deliberation process.} For all repair proposals
            $\candidateblk_j \in \rcbpool$ satisfying $\Blk'.\checkApproval(\policy,\vote_j) = \voting$ (where $\vote_j$ is vote for $\candidateblk_j$), that the node is willing to endorse,   
            \begin{compactenum}
            	\item Parse the proposal $\candidateblk_j := \langle\parsex{j}{\star}\rangle$
            	\item Generate $v_j \gets \Blk'.\votefunc(C,s_j^\star)$. Create a vote transaction $\voteTx$ with $\voteTx.$\texttt{data}$= \vote_j$
            	\item Broadcast $\voteTx$. 
            \end{compactenum}
\end{flushleft}	
	
\begin{flushleft}
\textbf{Proposing a new block.} Collect all transactions, denoted by $\txset$ from the network for
the $\round$-th round and try to build a new block $B_\round$: 
    \begin{compactenum}
        \item \emph{(Determine state transition from the head of the chain).}
            Repair the chain by applying the repair proposals that are
            approved: $\forall \candidateblk_j = \parsex{j}{\star}$ such that
            $\Blk'.\checkApproval(\policy, \vote_j) =
            \accept$, where $\vote_j \gets \Blk'.\votefunc(\chain,\candidateblk_j)$, set $\accset = \delta(\accset_{r-1},\txset)$.
		\item \emph{(Proof of Stake).} Extend chain and $\layer$ data structures as follows, 
		\begin{compactenum}
			\item Let $\account^*$ be the account of the stakeholder, generate $\sigma \gets \solvepos(\chain,\account^*,(H(\header_{r-1}),G(\parsex{}{})))$
			\item Set $\hdata := (\sigma,\account^*)$ and complete $\header$ by appropriately setting other values
			\item Set new block $B_r \gets \langle \header, \parsex{}{} \rangle$
			\item Extend local chain $\chain \gets \chain \vert\vert B_r$, the repair layer $\Tail \gets \Tail \vert\vert \Tail_{r}$ and the approved repairs $\edit \gets \edit \vert \vert \edit_{r}$
			\item Then broadcast $(\chain,\Tail,\edit)$ to the network
		\end{compactenum}
    \end{compactenum}
\end{flushleft}

\begin{flushleft}
    \textbf{Updating the chain.} When a node receives $\chain,
    \Tail,\text{and}\ \edit$, check if the chain is valid by calling
    $\Blk'.\chainValid(\chain, \Tail, \edit) = 1$. Accept the new chain if the new chain is valid as
    per PoS's fork resolution rule.
\end{flushleft}

\caption{$\layer$ protocol integration into Ethereum with Algorand or Ouroboros Praos as the underlying consensus and parameterized by policy $\policy$. Meaning $\solvepos$ is instantiated with Algorand or Praos.}
\label{fig:ethprotocol_pos}
\end{figure*}

\subsection{Reparo in Ethereum with PoW}\label{sec:ethereum}

We present here how one would instantiate $\layer$ in Ethereum with PoW based consensus done currently. We present here only the differences from the PoS instance we presented in~\cref{sec:ethereum_pos}.

\floatstyle{boxed}
\restylefloat{figure}

\begin{figure*}[!htb]
\smaller
The Ethereum protocol consists of a sequence of rounds $\round$. 

\begin{flushleft}
\textbf{Initialization.} We initialize new databases the repair layer
$\Tail_0 \leftarrow \genesis$, and the approved repairs $\edit_0 \gets \emptyset$, set
round $\round \leftarrow 1$ and an empty list of repair proposals $\rcbpool \gets \emptyset$.
\end{flushleft}
For each round $\round$, first initialize $\edit_{r}
\gets \emptyset$, $\Tail_{r} \gets \emptyset$ and we describe the following
sequence of execution.

\begin{flushleft}
\textbf{Proposal.} A node creates a repair proposal $\candidateblk_j
\leftarrow \Blk'.\redactreq(\chain,j, \txset_j^\star)$ (refer~\cref{alg:request_pos}) for block $B_j, j \in [r-1]$ using transactions
$\txset_j^\star$. It then broadcasts it to the network.
\end{flushleft}

\begin{flushleft}
\textbf{Update Proposal pool.} Collect all repair proposals $\candidateblk_j$
from the network and add $\candidateblk_j$ to $\rcbpool$ iff $\candidateblk_j$
is valid; otherwise discard $\candidateblk_j$.
\end{flushleft}

\begin{flushleft}		
\textbf{Repairing the chain.} For all repair proposals $\candidateblk_j :=
\langle \parsex{j}{\star} \rangle \in \rcbpool$, we denote a vote $\vote_j \gets \Blk'.\votefunc(\chain,\candidateblk_j)$ and do:
	\begin{compactenum}
        \item If $\Blk'.\checkApproval(\policy, \vote_j) = \accept$,
            then call algorithm $(\chain',\Tail')\gets \Blk'.\editchain(\chain,\Tail,\candidateblk_j)$. Here $j$-th block in $\chain'$ is $\langle
            \header_j,\candidateblk_j \rangle$ and subsequent blocks' states
            are updated accordingly. Then do the following,
         \begin{enumerate}
         	\item Add $\txset_j^\star$ to $\edit_{r}$ and remove $\candidateblk_j$ from $\rcbpool$
         	\item set local chain $\chain = \chain'$ and update $\Tail = \Tail'$
         \end{enumerate}
        \item If $\Blk'.\checkApproval(\policy, \vote_j) =
            \reject$, then remove $\candidateblk_j$ from $\rcbpool$ 
        \item If $\Blk'.\checkApproval(\policy, \vote_j) = \voting$, then do nothing 
	\end{compactenum}
\end{flushleft}
	
\begin{flushleft}
\textbf{Mining a new block.} Collect all transactions, denoted by $\txset$ from the network for
the $\round$-th round and try to build a new block $B_\round$: 
    \begin{compactenum}
        \item \label{step:deliberation}\emph{(Deliberation process).} For all repair proposals
            $\candidateblk_j \in \rcbpool$ that the node is willing to endorse,   
            \begin{compactenum}
            	\item Parse the proposal $\candidateblk_j := \langle\parsex{j}{\star}\rangle$
            	\item Generate $v_j \gets \Blk'.\votefunc(C,s_j^\star)$. If $\Blk'.\checkApproval(\policy, \vote_j) = \voting$ then create a vote transaction $\voteTx$ with $\voteTx.$\texttt{data}$= \vote_j$
            	\item Update $\txset \leftarrow \txset \vert\vert \voteTx$
            \end{compactenum}
        \item \emph{(Determine state transition from the head of the chain).}
            Repair the chain by applying the repair proposals that are
            approved: $\forall \candidateblk_j = \parsex{j}{\star}$ such that
            $\Blk'.\checkApproval(\policy, \vote_j) =
            \accept$, where $\vote_j \gets \Blk'.\votefunc(\chain,\candidateblk_j)$, set $\accset = \delta(\accset_{r-1},\txset)$.
		\item \emph{(Mining).} Extend chain and $\layer$ data structures as follows, 
		\begin{compactenum}
			\item Perform standard Ethereum mining and set new block $B_r \gets \langle \header, \parsex{}{} \rangle$
			\item Extend local chain $\chain \gets \chain \vert\vert B_r$, the repair layer $\Tail \gets \Tail \vert\vert \Tail_{r}$ and the approved repairs $\edit \gets \edit \vert \vert \edit_{r}$
			\item Then broadcast $(\chain,\Tail,\edit)$ to the network
		\end{compactenum}
    \end{compactenum}
\end{flushleft}

\begin{flushleft}
    \textbf{Updating the chain.} When a node receives $\chain,
    \Tail,\text{and}\ \edit$, check if the chain is valid by calling
    $\Blk'.\chainValid(\chain, \Tail, \edit) = 1$. Accept the new chain if the new chain is valid as
    per Ethereum's fork resolution rule\footnotemark.
\end{flushleft}

\caption{$\layer$ protocol integration into Ethereum with PoW based consensus and parameterized by policy $\policy$}
\label{fig:ethprotocol}
\end{figure*}
\footnotetext{Ethereum uses Greedy Heaviest
    Order Sub Tree (GHOST) protocol to rank chains and this is slightly
different from the longest chain rule used in Bitcoin.}

\paragraph{Repair Policy} 
A repair proposal is approved according to $\Blk'.\checkApproval$ with policy $\policy$, if the following conditions hold:
\begin{asparaitem}

\item The proposal does not propose to modify the address fields or the value
    field of a transaction.

\item The proposal is unambiguously not a double spend attack attempt (needs 
    information from the real world for confirmation).
\item The proposal does not redact or modify votes in the chain.
\item The proposal has received more than $\rho$ fraction of votes ($50$\% of
    votes) in $\ell$ consecutive blocks\footnote{The probability that a
      malicious proposal is accepted by \layer{} is always
      $< \ell\widetilde{\rho}^{(\tfrac{\ell}{2}+1)}$, where $\widetilde{\rho} < \tfrac{1}{2}$ is the
      fraction of Byzantine miners. This is negligible for a sufficiently large
      choice of $\ell$.} (voting period, that can decided by the
    system) after the corresponding $\redactTx$ is included in the chain.
\end{asparaitem}

%
%

\paragraph{Performing Repair Operations} Upon approval with respect to the policy $\policy$ repair operations are performed as they were performed in the PoS variant. Additionally, in the PoW system, there are no restrictions on the redaction policy except for the ones discussed above. Unlike the PoS variant, we can redact transactions as a whole since the consensus is independent of the state of accounts. 

\pgfarrowsdeclarecombine{twotriang}{twotriang}
{stealth'}{stealth'}{stealth'}{stealth'}

\paragraph{Block Validation} Block validation is the same as in the PoS variant, except that now instead of checking for PoS consensus we check if the block has the correct nonce for PoW. A formal description of the  procedure can be found
in~\cref{alg:validateblk}. 

\begin{algorithm}[b]
\small
\SetAlgoVlined
\SetKwInOut{Input}{input}\SetKwInOut{Output}{output}
\Input{Chain $\chain = (B_1,\cdots,B_n)$, repair layer $\Tail=
(\Tail_{1},\cdots,\Tail_{n})$, block $B_{n+1}$, repair approved $\edit_{n+1}$.}
\Output{$\{\bot,(\chain',\Tail')\}$}

\BlankLine
Parse $B_{n+1} := \generalblocki{n+1}$, where $\header_{n+1} =
(\parentheader_{n+1}, G(\parsex{n+1}{}),\hdata_{n+1})$\;
Parse $B_n := \generalblocki{n}$, where $\header_n = (\parentheader_n, G(\parsex{n}{}),\hdata_n)$\;
Validate transactions $x_{n+1}$, if \emph{invalid} \Return $\bot$\;
\lIf{$\parentheader_{n+1} \ne H(\header_n)$}{\Return $\bot$}
\lIf{$\edit_{n+1} = \emptyset \land \accset_{n+1} = \transition(y_n,x_{n+1}) \land
\allowbreak \checkpow(\header_{n+1})$}{ Set $\chain' \gets \chain \vert\vert B_{n+1}$, and $\Tail' \gets \Tail \vert \vert
\emptyset$, and \Return $(\chain',\Tail')$}
\Comment{\scriptsize{Validate Blocks where repairs were approved}}
Initialize $\chain' \gets \chain, \Tail' \gets \Tail$\;
\For{all $\txset_j^\star \in \edit_{n+1} $}{
\Comment{\scriptsize{Perform all the repair operations}}
    Parse $B_j := \generalblocki{j}$, where  $\header_j = (\parentheader_j,
    G(\parsex{j}{'}),\hdata_j)$\;
    \lIf{$\checkApproval\left(\policy, H\left(\Gtxroot(\txset_j),\Gtxroot(\txset_j^\star)\right)\right) \ne
    \accept$}{\Return $\bot$}
    Set $\accset_j^\star := \transition(\accset_{j-1},\txset_j^\star)$\;
    \Comment{\scriptsize{Perform the repairs as originally performed}}
    $\chain', \Tail' \gets \editchain(\chain', \Tail',
    \candidateblk_j )$\;
}
Parse $\chain' := (B_1', \cdots, B_n')$ and $B_n' := \langle \header'_n, \parsex{n}{'} \rangle$\;
\Comment{\scriptsize{Check the state transition after repair}}
\lIf{$\accset_{n+1} = \transition(\accset'_n, \txset_{n+1}) \land \allowbreak 
\checkpow(\header_{n+1})$}{Set $\chain' \gets \chain \vert\vert B_{n+1}$, and $\Tail' \gets \Tail \vert \vert
\emptyset$ and \Return $(\chain',\Tail')$} 
\Return $\bot$\;
\caption{$\blockValid$ }	
\label{alg:validateblk}
\end{algorithm}

Security and public verifiability properties are the same way as in the PoS variant. We describe the security argument here for completeness.

\paragraph{Security} Since the hash function $H$ is modeled as a random oracle (RO), finding a collision on a vote (which is the hash of the ID of the old transaction and the ID of the candidate transaction) is highly improbable. Therefore, when a miner votes for a repair proposal in his newly mined block, no adversary can claim a different repair proposal for the same vote value. Same property of the hash function $H$ also ensures that no adversary can find a different block that hashes to the same hash of an honestly mined block. Therefore an adversary cannot break the integrity of the chain. Together, they imply the \emph{unforgeability} of votes, as if an adversary wishes to vote, he has to mine a block with his vote himself. Assuming majority of the miners are honest in Ethereum, $\layer$ integration with Ethereum satisfies \emph{editable common prefix} and preserves chain quality and chain growth with respect to repair policy $\policy$.

\paragraph{Optimizations} \emph{State Assertion}: Instead of recomputing the
state, a $\editTx$ can propose a state for the affected contract and accounts.
The protocol can then inject this state if allowed by the policy. This
method is inexpensive as it is without any cascading computation and is useful
for users who accidentally locked their funds~\cite{EIP156}.




\subsection{Prominent Bugs}\label{sec:eth repair}

\begin{table}[ht]
    \centering
    \def\arraystretch{1.2}
	\caption{Prominent Smart Contract Bugs}

    \begin{tabular}{l rr}
      \toprule
		\textbf{Bug} & \textbf{Start Block} & \textbf{ETH Affected}
		\\ \midrule
		DAO\footnotemark & $1,428,757$ & $3,600,000$ \\ 
		QuadrigaCX\footnotemark & $1,952,428$ & $67,317$ \\
        Parity Multisig\footnotemark & $4,049,249$ & $513,736$ \\ 
		REXmls\footnotemark & $4,066,859$ & $6,687$ \\ 
		No Code Contracts & $-$ & $6533.17$ \\
    \bottomrule
    \end{tabular}
    \label{table:bug info}
\end{table}

\addtocounter{footnote}{-3}
\footnotetext{DAO with transaction hash: \texttt{0xe9ebfecc2fa10100db51a} -
\texttt{4408d18193b3ac504584b51a4e55bdef1318f0a30f9}}
\stepcounter{footnote}
\footnotetext{Quadriga CX with transaction hash:
\texttt{0xf4c8423215e8abb2810ff} - \texttt{0b2eb82a93ce660a0ce651df5f14005e08f6e25318e}}
\stepcounter{footnote}
\footnotetext{Parity with transaction hash:
\texttt{0x348ec4b5a396c95b4a5524ab0} - \texttt{ff61b5f6e434098cf6e5c1a6887bed2bc35625d}}
\stepcounter{footnote}
\footnotetext{REXmls (imbrex) token with transaction hash:
\texttt{0xcb6b1e83452608} - \texttt{65b5e164e09584634960ec1d02d80d3bbb4e1533c77393d216}}
\stepcounter{footnote}

In \cref{table:bug info} we provide the information of some of the popular bugs.

\paragraph{DAO} This is a re-entrancy bug in the contract that allowed a
maliciously crafted call to drain the balance of the contract before it
subtracted the balance from the user. We propose to fix this contract by
updating all DAO contract creation contracts with the bug fixed code. This is
different from the ad-hoc solution employed by Ethereum today. Ethereum
hard-coded the address for DAO and executes the contract differently. This
ensured that the blockchain should have no transaction dependency because the
blockchain already has the state with the contract fixed. This in conjunction
with the repair proposal allows an inexpensive repair for DAO even though it has
a lot of dependent transactions.

\paragraph{Parity Multisig Wallet Bug} The Parity Mutli Sig Wallet is a library
contract that had a bug which had a public constructor that allowed any user to
take control of the contract. A user took ownership of the contract and
accidentally killed it. We propose to repair this contract by undoing the
transaction that killed the contract. The transaction dependency is unaffected
as it just resurrects a dead contract. This enables all Parity Multisig Wallet
holders to safely recover their funds.

\paragraph{QuadrigaCX (QCX) and REXmls} These contracts have hardcoded wrong
addresses in the contract which sent the ICO ETH to an incorrect address (an
account that does not exist) thereby permanently locking the coins in those
contracts. We propose to repair this bug by proposing a repair transaction with
the same code but with the correct address, which can be used to recover and
return the lost funds.

\paragraph{No code contract} There are $2,986$ such contract creation
transactions which have money but no code in the creation call.
The idea to solve the no code contract problem, is to allow the user to add code
to the contract. We give a template of the code in \cref{fig:contract no code
fix}. The contract allows the user who locked the money in
a contract to retrieve the money.
\begin{figure}
\begin{verbatim}
  pragma solidity ^0.5.0;
  
  contract simpleWithdraw {
      address private owner;
      uint256 money;
      constructor() public payable {
          owner = msg.sender;
          money = msg.value;
      }
  
      function withDraw() public {
          if (msg.sender == owner) {
              selfdestruct(msg.sender);
          }
      }
  }
\end{verbatim}
\caption{Contract to retrieve lost money}
\label{fig:contract no code fix}
\end{figure}

\section{Other Instantiations}\label{sec:other}
In this section we discuss how our protocol can be instantiated into other systems like Bitcoin and Cardano. Note that Cardano is a Proof of Stake~\cite{kiayias2017ouroboros,CCS:BGKRZ18} based system. Even though Bitcoin and Ethereum are PoW based systems, we discuss Bitcoin instantiation because redaction of illicit data entries is a major problem in Bitcoin and we want to highlight our improvements compared to the work of Deuber et al.~\cite{deuber2019redactable}. 

\subsection{Integrating into Bitcoin}\label{subsec:bitcoin} Deuber et al.~\cite{deuber2019redactable} instantiated their redactable blockchain protocol with Bitcoin and showed how to redact harmful illicit content from data pockets in transactions. We now describe how to instantiate our $\layer$ protocol from~\cref{fig:newprotocol} on top of Bitcoin for removal of arbitrary non-payment data bytes (\emph{stateless} repairs) and any other repair operations if such a need arises. We primarily focus on the former repair operations in case Bitcoin as removal of illicit data entries is immediate critical problem to solve.

The main differences between~\cite{deuber2019redactable} and ours when $\layer$ is integrated into Bitcoin are: 
\begin{enumerate}
	\item  Our protocol does not require the modification of the Bitcoin block header.
	\item As a consequence of the above point, we do not require an additional hash value to be stored in every Bitcoin block header like their protocol. This makes our protocol much more space efficient than theirs. 
	\item Our protocol is immediately integrable into Bitcoin in a backward compatible fashion. This means that already existing illicit data entries in Bitcoin can be redacted once $\layer$ is fit on top of Bitcoin today.
	\item Unlike~\cite{deuber2019redactable}, we handle both stateless redactions and stateful repair operations.
\end{enumerate}

We refer to~\cite{deuber2019redactable} for basic understanding of how Bitcoin transactions work and how arbitrary data can be inserted into transactions.

Regarding user roles, as described in~\cref{sec:ethereum_pos}, \emph{miners} also happen to be \emph{deciders} in the instance of Bitcoin. \emph{Miners} vote for  proposal s as they happen to mine blocks similar to the protocol in~\cite{deuber2019redactable}.

\paragraph{Data Structures} Note that unlike~\cite{deuber2019redactable} we do not require any modification of the Bitcoin block structure. Instead we have a \emph{repair layer} $\Tail$ for each block that is empty when the block is mined. 

\paragraph{Proposing Repairs} Similar to~\cite{deuber2019redactable} we have a special transaction $\redactTx$ that contains the hash of the old and new version of the transaction in its output script. In other words the transaction id's $\tx_{\id}$ and $\tx_{\id}^\star$ are stored. In this case $\tx^\star$ is the candidate transaction. The user then broadcasts  $\redactTx$ and $\tx^\star$ to the network; $\redactTx$ requires a transaction fee to be included in the blockchain, while $\tx^\star$ is added to a pool of candidate transactions. The candidate transaction $\tx^\star$ is validated by checking its contents with respect to $\tx$, and if it is valid, then it can be considered for voting. For stateless repairs like redacting arbitrary non-payment data entries only, it is checked if the only difference between $\tx$ and $\tx^\star$ is that of the missing data entry in $\tx^\star$.

\paragraph{Repair Policy} Our protocol is parameterized by a repair policy parameter $\policy$ similar to~\cite{deuber2019redactable}. For the case of redacting non-payment data entries we follow the same basic policy recommendations as theirs. For completeness, we detail the policy requirements here. A proposed redaction is approved valid if the following conditions hold:
 \begin{itemize}
  	\item It is identical to the transaction being replaced, except that it can remove data.

 	\item It can only remove data that can never be spent, e.g.,\ $\mathtt{OP\_RETURN}$ output scripts.
     
    \item It does not redact votes for other redactions in the chain. 
 	
 	\item It received more than $50$\% of votes in the $1024$ consecutive blocks (voting period) after the corresponding $\redactTx$ is stable in the chain.

 \end{itemize}

Similar to the discussion on repair policy for Ethereum~\cref{sec:ethereum_pos}, we argue that the policy for handling such stateful repair  proposal s can be quite event specific and strongly dependent on auxiliary information from the real world. The auxiliary information from the real world helps to see if a particular stateful repair  proposal  is good for the chain or not. As a minimum requirement from the policy (to help miners detect malicious proposals), a proposed stateful repair operation is approved valid if the following conditions hold:
 \begin{itemize}
  	\item It is unambiguosly not a double spend attack attempt (needs auxiliary information from the real world for confirmation).

 	\item It cannot propose to repair a transaction thereby making it an invalid spend.
 	
 	\item It does not change the amounts being transacted.
     
    \item It does not redact or modify votes in the chain. 
 	
 	\item It received more than $50$\% of votes in the $1024$ consecutive blocks (voting period) after the corresponding $\redactTx$ is stable in the chain.

 \end{itemize}

Deliberation for a proposal is done via voting by miners. Voting for a candidate transaction $\tx^\star$ simply means that the miner includes $\redactTx_\id = H(\tx_{\id} || \tx_{\id}^\star)$ in the coinbase (transaction) of the new block he produces. After the voting phase is over as determined by the policy $\policy$, the candidate transaction is removed from the candidate pool. 

\paragraph{Performing Repair Operations} Our protocol slightly varies from~\cite{deuber2019redactable} in this regard. In case of stateless repair operations like redactions, once a candidate transaction has been approved by the redaction policy, the miners in the network replace the old version of the transaction (being repaired) and replaces it with the candidate transaction, while storing the hash of the old version in the corresponding repair layer $\Tail$ of the block. In case of stateful repair operations, the entire old version of the transaction is stored in the corresponding repair layer of the block. 

\paragraph{Updating UTXO with Stateful Repair Operations} Since stateful repair operations could involve changing payment information which could affect the UTXO database after the mining of a block, special care needs to be taken in updating the UTXO. When a repair operation  proposal  is approved, the miners perform the repair operation as discussed above. The UTXOs after each of the following blocks upto the most recent block is accordingly updated. This could mean that some transactions in these blocks become invalid spends as their input is no longer in the UTXO database at the time that block was mined (after performing the repair operation). This is the cascading state update that was discussed previously in the case of Ethereum~\cref{sec:ethereum}. After performing the repair operation and reflecting it in the UTXO database at each subsequent block, the miners try to mine a new block by including transactions that are consistent with the updated UTXO database at that time.

 \paragraph{Chain Validation} To validate a full chain a miner needs to validate all the blocks within the chain. We discuss the general case where both stateless (redactions) and stateful repair operations could have been performed on the chain. Consequently our chain validation procedure is different from~\cite{deuber2019redactable} as ours is more generic. For validation, the miner uses the repair layer of the blocks in the chain to \emph{go back in time} to the mined version of the blocks. This ensures that the miner is now having the the original state of the chain. Note that this holds true even in case of stateless redactions as the repair layer $\Tail$ would contain the hash of old version of the transaction. The miner now validates the chain from the genesis by ``re-mining" the chain, with the catch that instead of solving for PoW, he verifies the existing PoW. Since the miner is validating the blocks as if when they were freshly mined, a valid PoW (in the past) remains a valid PoW now for the miner. If ever some stateful or stateless repair proposal  was approved, the miner performs the repair operation that is required to be performed (as it would have been performed). This way the miner validates and re-constructs the chain. The miner rejects a chain as invalid if \emph{any} of the following holds: (1) a block's repair operation was not approved according to the policy, or (2) a previously approved repair was not performed on the chain.

\paragraph{Validating Transactions} Validating a chain involves validating a  block and its contents as a subroutine. The miner validates all the transactions contained in its transactions list against the current database of UTXOs; the validation of unedited transactions is performed in the same way as in the immutable version of the Bitcoin protocol. The miner simply validates transactions in the block against their witnesses. In case of having only the hash old version of the transaction and the old witness (this is the case of removal of non-payment data entries - stateless redactions), the miner can validate the witness with respect to the new version of the transaction as the payment scripts are unchanged in a stateless redaction. This is similar to the validation in~\cite{deuber2019redactable}. Therefore, we can ensure that all the transactions included in the block have a valid witness, or in case of redacted transactions, the old version of the transaction had a valid witness.

\subsection{Integrating into Cardano (PoS)}\label{subsec:cardano} Cardano is a cryptocurrency that runs the Ouroboros PoS consensus mechanism. Our interest in this system is to show how one can instantiate our $\layer$  protocol~\cref{fig:newprotocol} on top of a PoS based blockchain. 

\paragraph{Similarities between Bitcoin and Cardano} Transactions in Cardano work the same way as in Bitcoin and Cardano is based on a UTXO model. The address field in Cardano has additional semantics for the staking procedure of the PoS consensus process. Block headers in Cardano are more or less the same as in Bitcoin except for consensus proof which is different from the PoW value in Bitcoin.

\paragraph{Consensus} The time is divided into 120 second slots and each slot has a slot leader elected to propose the new block. The slot leaders are elected with a winning probability proportional to their stake in the system. In an epoch which lasts for 20 hours, the slot leaders for each slot of the next epoch are determined but not revealed. As a proof of election, the elected slot leader generates a signature proving his stake in the system, which can be verified by everyone else.

\paragraph{Policy for PoS} The repair policy requirements is more or less the same as discussed previously for systems like Ethereum and Bitcoin. However, now we do not allow \emph{redactions} that affect the state of the system. In case of a \emph{stateful} redaction, as the state of the chain has changed, the data point that causes this state change is erased/redacted. In PoS based blockchain systems, such a stateful redaction causes failure in chain validation as an honest new user can no longer verify the stakeholder consensus proof. This is because the stake distribution has changed, but the transaction resulting in the older stake distribution is no longer stored and therefore consensus proofs based on the older stake distribution can no longer be verified.

\paragraph{Repairing and Chain Validation} The only difference in terms of proposing and finally performing a repair operation is that the voting period for a repair proposal begins from start of the immediately next epoch and spans throughout that epoch. This is to ensure that chain validation procedure is able to validate consensus after repair operations have been performed. In more detail, recall that the chain validation as described for Bitcoin, we \emph{go back in time} and validate the blocks in the state in which they were mined and perform repair operations just the way they were performed. When following this procedure, one must be able to validate the consensus proof, PoW in Bitcoin and PoS in Cardano. As mentioned earlier, the slot leaders for the current epoch are determined by the end of the previous epoch. Consider a case where there is a repair proposal  whose voting period starts in the middle of an epoch and ends in the middle of the next epoch and gets approved. The miner performs the repair operation that could potentially change the state and thereby the stakes. This makes the verification of the elected slot leaders for the rest half of the slots in the epoch inefficient and time consuming, as these slot leaders were determined by the state before the repair operation was performed. In order to avoid this inefficiency, we let the voting period to be synchronized with the epoch period of Cardano. Chain validation can now proceed as in Bitcoin, except that the consensus is PoS~\cite{kiayias2017ouroboros}.

\fi

\end{document}